\documentclass[a4paper]{llncs}
\usepackage[english]{babel}
\usepackage{german}
\usepackage{amssymb}
\usepackage{amsfonts}
\usepackage{amsmath}
\usepackage[latin1]{inputenc}
\usepackage{fullpage}
\usepackage{cite}
\usepackage{graphicx}
\usepackage{subfigure}
\usepackage{algorithm}
\usepackage[noend]{algorithmic}
\usepackage{amsmath,amssymb,cite}
\usepackage{footmisc,marvosym}
\usepackage{wasysym,stackrel}

\newtheorem{observation}{Observation}

\begin{document}
\title{Determining Majority in Networks with Local\\ Interactions and very Small Local Memory \thanks{A preliminary conference version of this work will appear in the \emph{41st International Colloquium on Automata, Languages, and Programming (ICALP)}, 
Copenhagen, Denmark, 2014.}} 
\author{George B. Mertzios\inst{1}\thanks{Partially supported by the EPSRC Grant~EP/K022660/1.} 
\and Sotiris E. Nikoletseas\inst{2,3}\thanks{Partially supported by the MULTIPLEX project~--~317532.}
\and \\
Christoforos L. Raptopoulos\inst{2,4}\thanks{Partially supported by the SHARPEN project~--~PE6 (1081).} 
\and Paul G. Spirakis\inst{2,3,5}\thanks{Partially supported by the MULTIPLEX project~--~317532, the EU ERC Project ALGAME and the EEE/CS School of the University of Liverpool.}}

\institute{School of Engineering and Computing Sciences, Durham University, UK\\
\and Computer Technology Institute and Press ``Diophantus'', Greece\\
\and Computer Schience Department, University of Patras, Greece \\
\and Computer Science Department, University of Geneva, Switzerland\\
\and Department of Computer Science, University of Liverpool, UK\\
\email{george.mertzios@durham.ac.uk}, \email{nikole@cti.gr}, \\
\email{raptopox@ceid.upatras.gr}, \email{p.spirakis@liverpool.ac.uk}}

\date{\vspace{-1cm}}
\maketitle

\begin{abstract}
We study here the problem of determining the majority type in an arbitrary connected
network, each vertex of which has initially two possible types (states). 
The vertices may have a few additional possible states and can interact in pairs only if
they share an edge. Any (population) protocol is required to
stabilize in the initial majority, i.e. its output function must interpret the local state
of each vertex so that each vertex outputs the initial majority type. We first provide a protocol with 4 states per vertex that \emph{always} computes the initial majority value, 
under any fair scheduler. 
Under the uniform probabilistic scheduler of pairwise interactions, 
we prove that our protocol stabilizes in expected polynomial time for any network and is quite fast on the clique.
As we prove, this protocol is optimal, in the sense that there does
not exist any population protocol that always computes majority with fewer than 4 states per vertex.
However this does not rule out the existence of a protocol with 3 states per vertex that is correct with high probability (whp). 
To this end, we examine an elegant and very natural majority protocol with 3 states per vertex, introduced in \cite{AAE08} 
where its performance has been analyzed for the clique graph. 
In particular, it determines the correct initial majority type in the clique very fast and whp under the uniform probabilistic scheduler. 
We study the performance of this protocol in arbitrary networks. 
We prove that, when the two initial states are put uniformly at random on the vertices, the protocol of \cite{AAE08} converges to the initial majority 
with probability higher than the probability of converging to the initial minority. 
In contrast, we present an infinite family of graphs, on which the protocol of \cite{AAE08} can fail, i.e. it can converge to 
the initial minority type whp, even when the difference between the initial majority and the initial minority is $n - \Theta(\ln{n})$. 
We also present another infinite family of graphs in which the protocol of \cite{AAE08} takes an expected exponential time to converge. 
These two negative results build upon a very positive result concerning the robustness of the protocol of \cite{AAE08} on the clique, 
namely that if the initial minority~is~at~most~$\frac{n}{7}$, the protocol fails with exponentially small probability. 
Surprisingly, the 
resistance of the clique to failure 
causes the failure 
in general graphs. 
Our techniques use new domination and coupling arguments for suitably defined processes whose
dynamics capture the 
antagonism between the states involved.
\end{abstract}

\section{Introduction\label{introduction}}

One of the most natural computational problems in many physical systems is
to compute the \emph{majority}, i.e.~to determine accurately which type of
an element of the system appears more frequently. For instance, the majority
problem is encountered in various settings such as in voting~\cite{HL75,KT08}, in
epidemiology and interacting particles systems~\cite{Lbook}, in diagnosis of multiprocessor systems~\cite{PMC67}, in
social networks~\cite{MNT14,Mmaster} etc. In distributed computing, the majority problem
is an important and natural special case of the central problem of reaching 
\emph{consencus} within a system ~\cite{LRP82,D74}, where a number of processes have to agree on any single data value
(e.g. leader election~\cite{FJ06}). In
all these physical systems, some pairs of elements may interact with each
other while other pairs may not be able to interact directly. This structure of the possible pairwise interactions between elements
of the system can be modeled by a network (i.e.~graph), where elements and
possible interactions are represented by vertices and edges, respectively.

In order to solve the majority computation problem in a network, we first
need to make some assumptions on the underlying model of computation. Much research has been done under the assumption that there exists a central authority, as well as unlimited available memory and full information about the whole network (see e.g.\cite{SW91,GA06}).
However, in many real systems we do not have (or
we do not wish to have) such a powerful computational model. 
The weaker the considered model of
computation is (e.g. lack of central authority, partial or no information about
the system, lack of memory etc.), the more challenging the majority
computation becomes.

One of the ways to study distributed systems where agents may interact in
pairs and each individual agent is extremely limited (in fact, being equipped
only with a finite number of possible states) is by using \emph{population protocols}
\cite{AADFP06,AR09}. Then the complex behavior of the system emerges from the rules
governing the possible pairwise interactions of the agents. Population
protocols have been defined by analogy to population processes~\cite{K87} in probability theory and have already been used in various fields, such as
in statistical physics, genetics, epidemiology, chemistry and biology~\cite{CSWB09}. 

In particular, population protocols are \emph{scalable}, i.e.~they work independently of the
size $n$ of the underlying network (called the \emph{interaction graph}) and
the value of $n$ is not even known to the protocol. Furthermore they are 
\emph{anonymous}, i.e.~there is only one transition function which is common
to all entities/agents: the result of an interaction of an agent $u$ at state $q_{u}$ with an agent $v$ at state $q_{v}$ is the same regardless of the identity of $u$ and $v$. The transition function of a population
protocol only specifies the result of every possible interaction, without
specifying \emph{which} pairs of agents interact or \emph{when} they are
chosen to interact. Usually it is assumed that interactions between agents
happen under some kind of a \emph{fairness} condition. For a survey about
population protocols we refer to~\cite{AR09}.

In this direction, a very natural and simple population protocol for the
majority problem on the clique (i.e.~the complete graph), where initially
every vertex has one of two possible types (states), has been introduced and
analyzed in~\cite{AAE08}. In particular, the protocol of~\cite{AAE08} assigns
only $3$ possible states to every agent (i.e.~there is a $3\times 3$
transition table capturing all possible interactions) and the interactions
between agents are dictated by a \emph{probabilistic scheduler} (i.e.~all
pairs have the same probability to interact at any step). Every vertex has
an identity $v$, but it is unaware of the identity of any other
vertex, as well as of its own identity. Although the underlying interaction
graph in~\cite{AAE08} is assumed to be a clique, the authors distinguish in
their protocol the agents $u$ and $v$ participating in an interaction into
an ``initiator'' and a ``responder'' of the interaction (when agents $u$ and $v$
interact, each of them becomes initiator or responder with equal
probability). Their main result is that, if initially the difference between
the initial majority from the initial minority in the complete graph with $n$
vertices is $\omega (\sqrt{n}\log n)$, their protocol converges to the
correct initial majority value in $O(n\log n)$ time with high probability. 

Most works on population and majority dynamics so far considered only two entity types
(e.g. the voter model~\cite{HL75}, the Moran process~\cite{M58}). The analysis of
population dynamics with more than two types is challenging. As an example we refer to the model of~\cite{AAE08}, in which, although agents can have initially one of only two types (red and green), the protocol itself allows every agent to be in one among three different states (red, green and blank) at every subsequent time point. Even though this model is quite simple, it is very hard to be analyzed. Computing the majority with as few states as possible
in the more general case, where the interaction graph has an arbitrary
structure (as opposed to the complete graph that has been mainly considered
so far) remained an open problem.

\subsection{Our contribution}

In this paper we study the majority problem in an \emph{arbitrary}
underlying interaction graph $G$, where initially every vertex has two
possible states (red and green). We consider here the weakest and simplest
possible model of computation. In particular, we assume the existence of no
central authority and we allow every vertex of $G$ to have only a (small)
constant number of available types (or states). Although every vertex of $G$
has a unique identity, no vertex is aware of its own identity or the
identity of any other vertex. Furthermore, although only two adjacent
vertices can interact, vertices of $G$ do not even know to which other
vertices they are adjacent.

First, we focus on the problem of \emph{always} computing the correct
majority value in an \emph{arbitrary} (directed or undirected) interaction graph $G$, regardless of how
large the initial difference between the majority and the minority is. In
particular, assuming that the interacting pairs of vertices are chosen by an arbitrary fair
scheduler, we derive matching lower and upper bounds on the number of
available states, for which there exists a population protocol that always
computes the correct majority value. For the lower bound, we prove that there does
not exist any population protocol that achieves this with at most $3$
different states per vertex. On the other hand, for the matching upper bound
we provide a population protocol with $4$ states per vertex, which always
computes the correct majority value, even if initially the difference
between majority and minority is $1$. To the best of our knowledge, this is the first 4-state population protocol that correctly computes the majority value in a two type population on an arbitrary interaction graph. In particular, the 4-state majority protocol proposed in~\cite{AR09} only works when the interaction graph is complete. Furthermore we provide polynomial
upper bounds on the expected time needed by our new protocol to converge,
and we show that in certain cases the running time is $O(n\log n)$, i.e.~the
same as for the fast protocol of~\cite{AAE08}.

Second, we provide a detailed analysis of the $3$-state protocol of~\cite{AAE08} on an arbitrary interaction graph $G$. Our first result in this
direction is that, when the two initial types (red and green) are
distributed on the vertices of an arbitrary graph $G$ uniformly at random,
the protocol of~\cite{AAE08} will converge to the initial majority with
higher probability than to the initial minority. The proof of this relies on a well known 
result in extremal combinatorics (in particular, on Hall's marriage
Theorem). Furthermore we present an infinite family of graphs $\{G_{n}\}_{n\in \mathbb{N}}$ on which the protocol of~\cite{AAE08} can fail (i.e.~it can converge to
the initial minority) with high probability, even when the difference
between the initial majority and the initial minority is as large as $n-\Theta(\ln{n})$. Then we
present another infinite family of graphs $\{G_{n}^{\prime }\}_{n\in \mathbb{N}}$ on which the protocol of~\cite{AAE08} can take an exponential expected number of
steps to converge. In particular, this rules out the possibility to use a
Markov chain Monte-Carlo approach to approximate the probability that the
protocol of~\cite{AAE08} converges to the correct majority value.

In order to prove our results on the classes $\{G_{n}\}_{n\in \mathbb{N}}$ and $\{G_{n}^{\prime }\}_{n\in \mathbb{N}}$, we first proved the intermediate result that for any $\varepsilon >0$,
if the minority has size at most $(\frac{1}{7}-\varepsilon )n$ in the
complete graph with $n$ vertices, then the protocol of~\cite{AAE08}
converges to the initial minority with exponentially small probability. The
latter result shows that, although the performance of the protocol of~\cite%
{AAE08} can drop significantly when the interaction graph $G$ is not the
complete graph, it is quite robust when $G$ is the complete graph. Our proof
concerning the robustness of the protocol of~\cite{AAE08} in the complete
graph is novel and uses a non-trivial coupling argument which can be of independent interest.

\section{The model and notation}

A \emph{population protocol} consists of a finite set $Q$ of states/types\footnote{In the original formulation of population protocols these are called \emph{states}, but we chose to also use the term \emph{type} in order to avoid confusion with the states in a Markov chain.}, a finite set of input symbols $X$, an input function $\iota :X\rightarrow Q$, a finite set of output symbols $Y$, an output function $\gamma: Q \to Y$, and a joint transition function $\delta: Q \times Q \to Q \times Q$. If, for any pair of states $q_1, q_2 \in Q$, $\delta(q_a, q_b) = (q'_a, q'_b)$ implies that $\delta(q_b, q_a) = (q'_b, q'_a)$, then the population protocol is called \emph{symmetric}. A population protocol is executed by a fixed finite \emph{population} of agents with types in $Q$. We assume that each agent has an identity $v \in V$, but agents are oblivious to their own identity and to identities of agents they interact with.

Initially, each agent is assigned a type according to an \emph{input} $x: V \to X$ that maps agent identities to input symbols. In the general population protocol model, agents are identified with the vertices of an \emph{interaction graph}, whose edges indicate the possible agent interactions that may take place. Here, an interaction graph is a simple connected graph $G$ (i.e.~~without loops or multiple edges), which can be directed or undirected. A function $C: V \to Q$ is called a \emph{configuration}. We will say that the population protocol reaches configuration $C$ at time $t$ if, for every agent $v$ in the population, the state of $v$ at time $t$ is $C(v)$.

In the original model, agents do not send messages or share memory; instead, an interaction between two agents updates both of their types according to a joint transition function (which can be also represented by a table). Interactions between agents are planned by a scheduler under a general ``fairness'' condition; the actual mechanism for choosing which agents interact is abstracted away. The fairness condition states that for any two configurations $C, C'$, if $C$ occurs infinitely often and $C'$ is reachable from $C$, then $C'$ also occurs infinitely often. 

In this paper, we consider a special case of a fair scheduler, namely the \emph{probabilistic scheduler}, which is defined on directed graphs as follows. During each execution step, a directed edge $(v, u) \in E$ is chosen uniformly at random from $E$, where $v$ (i.e.~the tail of $(v, u)$) is called the \emph{initiator} and $u$ (i.e.~the head of $(v, u)$) is called the \emph{responder} of the interaction. Then, agents $v$ and $u$ update their types jointly according to $\delta$. In particular, if $v$ is of type $q_v$ and $u$ is of type $q_u$, the type of $v$ (respectively $u$) becomes $q_v'$ (respectively $q_u'$), where $(q_v', q_u') = \delta(q_v, q_u)$. The types of all other agents remain unchanged. The probabilistic scheduler is defined on undirected graphs similarly, by replacing every undirected edge $\{v, u\}$ by the two directed edges $(v, u)$ and $(u, v)$. That is, in an undirected graph $G$, the probabilistic scheduler selects first an undirected edge uniformly at random and then it selects equiprobably one of its endpoint as the initiator.
Note that a symmetric protocol does not distinguish between initiators and responders. Therefore, if the protocol is symmetric, the probabilistic scheduler on undirected graphs just chooses at each execution step one undirected edge uniformly at random and lets its endpoints interact according to the transition function.

Given the probabilistic scheduler, a population protocol \emph{computes} a (possibly partial) function ${g: X^V \to Y}$ with error probability at most $\epsilon$, if for all $x \in g^{-1}(Y)$, the population eventually reaches a configuration $C$ that satisfies the following properties with probability at least $1-\epsilon$: (a) all agents agree on the correct output, i.e.~$g(x) = \gamma(C(v))$ for all $v \in V$ and (b) this is also true for every configuration reachable from $C$. 
Furthermore, a population protocol \emph{stably computes} a (possibly partial) function $g: X^V \to Y$ if, for \emph{every} fair scheduler, the population eventually reaches a configuration $C$ that satisfies both the above properties (a) and (b).

\begin{observation}
If a symmetric population protocol stably computes a function on an undirected interaction graph $G$, then it also stably computes the same function 
on a directed interaction graph $G'$ that comes from $G$ by assigning to every edge of $G$ one or two directions.
\end{observation}

\subsection{Majority with high probability on the clique} \label{AAE08-model}

Angluin et al.~\cite{AAE08} proposed a population protocol for computing majority with high probability (whp) in the case where the interaction graph is a clique. Their protocol uses just 3 types $Q = \{b, g, r\}$. For convenience, we will sometimes refer to these types as the \emph{blank, green} and \emph{red} type respectively. The joint transition function $\delta$ is given by:

\begin{equation}
\delta(x, y) = \left\{ 
\begin{array}{ll}
	(x, x), & \quad \textrm{if $x=y$ or $y = b$} \\
	(x, b), & \quad \textrm{if $(x, y) \in \{(g, r), (r, g)\}$}. 
\end{array}
\right.
\end{equation}

One of the main results in~\cite{AAE08} is that if the underlying interaction graph is a clique $K_n$ (i.e.~the complete graph on $n$ vertices) and interactions are planned according to the above probabilistic scheduler, then with high probability $1-o(1)$, the above 3-type majority protocol converges to the initial majority value if the difference between the initial majority and initial minority populations is $\omega(\sqrt{n} \log{n})$.

\subsection{Representation} \label{representation}

Using the probabilistic scheduler to plan agent interactions has the advantage that we can describe evolution by using a discrete time Markov chain ${\cal M}$. For general interaction graphs, the state space ${\cal S}$ of ${\cal M}$ can have up to $|V|^{|Q|}$ states, namely one for each configuration $C:V \to Q$. For two configurations $C, C'$, we will say that $C'$ is \emph{directly reachable from $C$} (or reachable in one step from $C$) if there is $(v, u) \in E$, such that $(C'(v), C'(u)) = \delta(C(v), C(u))$ and $C'(w) = C(w)$, for all $w \in V \backslash \{v, u\}$. 

Specifically for the model of~\cite{AAE08}, we denote by $W_t$ (respectively $R_t$ and $G_t$) the set of agents in state $b$ (respectively $r$ and $g$) at time $t$. Note that if the interaction graph has a high degree of symmetry, then ${\cal S}$ can be reduced significantly. One such example is the clique $K_n$, in which case we can describe a state of ${\cal M}$ by the tuple $(|R_t|, |G_t|)$ (where we have also used the fact that $|W_t| = n-|G_t|-|R_t|$).

\subsection{Preliminaries on Birth-Death Processes}

Consider a \emph{(discrete time) birth-death process} ${\cal B}$ with state space ${\cal S}_{\cal B} = \{S_0, S_1, \ldots, S_m\}$, for some integer $m \in \mathbb{N}$ and transition probability matrix $P$ given by 

\begin{equation}
P(S_i, S_j) = \left\{ 
\begin{array}{ll}
	p_i, & \quad \textrm{if $j=i+1$, for any $i=0, \ldots, m-1$} \\
	q_i, & \quad \textrm{if $j=i-1$, for any $i=1, \ldots, m$} \\
	1-p_i-q_i, & \quad \textrm{if $i=j \notin \{0, m\}$} \\
	1-p_i, & \quad \textrm{if $i=j=0$} \\
	1-q_i, & \quad \textrm{if $i=j=m$.} 
\end{array}
\right.
\end{equation}
In particular, when $p_0 = 1$ (resp. $p_0=0$), we say that ${\cal B}$ has a \emph{reflecting barrier} at state $S_0$ (resp. we say that $S_0$ is an \emph{absorbing state}). Similarly, when $q_m =1$ (resp. $q_m=0$), we say that ${\cal B}$ has a reflecting barrier at state $S_m$ (resp. we say that $S_m$ is an absorbing state). The fraction $\frac{p_i}{q_i}$ (resp. $\frac{q_i}{p_i}$) is called the \emph{forward bias} (resp. \emph{backward bias}) \emph{at state $S_i$}. If $\frac{p_i}{q_i} = \frac{p_j}{q_j}$, for all $i, j \neq \{0, m\}$, we refer to $\frac{p_i}{q_i}$ (resp. $\frac{q_i}{p_i}$) as the \emph{forward bias} (resp. \emph{backward bias}) \emph{of ${\cal B}$}.

We now present some useful preliminary results on discrete time birth-death processes. Other related results concerning birth-death processes can be found in~\cite{Nbook,Rbook}.

\begin{lemma}(Absorption probability) \label{Birth-death-absorption}
Consider a discrete time birth-death processes ${\cal B}$ satisfying $p_i=p$ and $q_i=q$, with $p \neq q$, for all $i=1, \ldots, m-1$. Then, given that the process starts at state $S_i$, the probability of reaching $S_0$ before reaching $S_m$ is equal to 

\begin{equation}
\Pr\{\textrm{${\cal B}$ reaches $S_m$ before $S_0$}| {\cal B}(0) = S_i\} = \frac{\left( \frac{q}{p}\right)^i -1}{\left( \frac{q}{p}\right)^m -1}.
\end{equation}
In particular, $\Pr\{\textrm{${\cal B}$ reaches $S_m$ before $S_0$}| {\cal B}(0) = S_1\} = \frac{\frac{q}{p} -1}{\left( \frac{q}{p}\right)^m -1}$.
\end{lemma}
\proof Let $h_i = \Pr\{\textrm{${\cal B}$ reaches $S_m$ before $S_0$}| {\cal B}(0) = S_i\}$, where $i=0, \ldots m$. Then $h_0=0, h_m=1$ and $h_i = ph_{i+1}+qh_{i-1} + (1-p-q)h_i$, for $i=1, \ldots, m-1$. Setting $u_i = h_i-h_{i-1}$, we then have that $u_{i+1} = \frac{q}{p} u_i$, hence $u_{i+1} = \left(\frac{q}{p}\right)^i u_1$, for $i=1, \ldots, m-1$. Furthermore, $1 = u_1+\cdots+u_m = \left( 1+ \frac{q}{p} + \cdots + \left(\frac{q}{p}\right)^{m-1}\right)u_1$. Since $u_1 = h_1$, we have that $h_1 = \frac{1}{1+ \frac{q}{p} + \cdots + \left(\frac{q}{p}\right)^{m-1}} = \frac{\frac{q}{p} -1}{\left( \frac{q}{p}\right)^m -1}$. Finally, we have that $h_i = \sum_{j=0}^{i-1} u_{j+1} = \sum_{j=0}^{i-1} \left(\frac{q}{p}\right)^j u_1 = \frac{1+ \frac{q}{p} + \cdots + \left(\frac{q}{p}\right)^{i-1}}{1+ \frac{q}{p} + \cdots + \left(\frac{q}{p}\right)^{m-1}}$, which completes the proof. \qed

\begin{lemma}(Expected time to absorption) \label{Birth-death-Time}
Consider a discrete time birth-death processes ${\cal B}$ satisfying $p_i=p$ and $q_i=q$, with $p \neq q$, for all $i=1, \ldots, m-1$. If ${\cal B}$ has a reflecting barrier at $S_0$ (i.e.~$p_0=1$), then the expected time that ${\cal B}$ takes to reach $S_m$ is 

\begin{equation}
\mathbb{E}[\textrm{time for ${\cal B}$ to reach $S_m$}| {\cal B}(0) = S_i] = \left( 1+ \frac{1}{p \left(\frac{q}{p}-1\right)}\right) \frac{\left( \frac{q}{p}\right)^{m} - \left( \frac{q}{p}\right)^{i}}{\frac{q}{p} - 1} - \frac{m-i}{p \left(\frac{q}{p}-1\right)}.
\end{equation}
In particular, for large enough $m$, we have the following: (a) If $\frac{q}{p}<1$ and $i<m$, then 

\begin{equation}
\mathbb{E}[\textrm{time for ${\cal B}$ to reach $S_m$}| {\cal B}(0) = S_i] = O\left( \frac{m-i}{p \left(1-\frac{q}{p}\right)}\right).
\end{equation}
(b) If $\frac{q}{p}>1$ and $i=1$, then 

\begin{equation}
\mathbb{E}[\textrm{time for ${\cal B}$ to reach $S_m$}| {\cal B}(0) = S_1] = \Omega\left( \frac{1}{\frac{q}{p}-1} \left( \frac{q}{p} \right)^m \right).
\end{equation}

\end{lemma}
\proof Let $\mu_i = \mathbb{E}[\textrm{time for ${\cal B}$ to reach $S_m$}| {\cal B}(0) = S_i]$, where $i=0, \ldots m$. Then $\mu_0=1+\mu_1$, $\mu_m=0$ and $\mu_i = 1+p \mu_{i+1}+q\mu_{i-1}+(1-p-q) \mu_i$, for $i=1, \ldots, m-1$. Denote $y_{i+1} = \mu_{i+1}-\mu_i$, and thus $y_{i+1} = - \frac{1}{p} + \frac{q}{p} y_i = - \left(\frac{q}{p}\right)^i - \frac{1}{p} \left( 1 + \frac{q}{p} + \cdots + \left( \frac{q}{p} \right)^{i-1} \right) = - \left(\frac{q}{p}\right)^i - \frac{1}{p}\frac{\left(\frac{q}{p}\right)^i -1}{\frac{q}{p}-1}$, for all $i=1, \ldots, m-1$. Note that $\sum_{i=1}^{m-1} y_{i+1} = -\mu_1$ and thus $\mu_1 = \sum_{i=1}^{m-1}\left[ \left( 1+ \frac{1}{p \left(\frac{q}{p}-1\right)}\right) \left( \frac{q}{p}\right) ^{i}- \frac{1}{p \left(\frac{q}{p}-1 \right)} \right] = \left( 1+\frac{1}{p \left(\frac{q}{p}-1 \right)}\right) \frac{\left( \frac{q}{p}\right)^{m}-\frac{q}{p}}{\frac{q}{p}-1}-\frac{m-1}{p \left(\frac{q}{p}-1 \right)}$. Furthermore, we have that 

\begin{eqnarray}
\mu_i - \mu_1 & = & \sum_{j=1}^{i-1} y_{j+1} = \sum_{j=1}^{i-1}\left[ -\left( 1+\frac{1}{p \left(\frac{q}{p}-1 \right)}\right) \left( \frac{q}{p}\right)^{j} + \frac{1}{p \left(\frac{q}{p}-1 \right)}\right] \\
& = & -\left( 1+\frac{1}{p \left(\frac{q}{p}-1 \right)}\right) \frac{\left( \frac{q}{p}\right)^{i}-\frac{q}{p}}{\frac{q}{p}-1} + \frac{i-1}{p \left(\frac{q}{p}-1 \right)}, \\
\end{eqnarray} 
and thus $\mu_i = \left( 1+\frac{1}{q-p}\right) \frac{\left( \frac{q}{p}\right)^{m} - \left( \frac{q}{p}\right)^{i}}{\frac{q}{p} - 1} - \frac{m-i}{p \left(\frac{q}{p}-1 \right)}$ as in the statement of the Lemma. Notice then that, if $\frac{q}{p}<1$ (or equivalently $q<p$) and $m$ is large enough, then $\mu_i$ is dominated by the term $\frac{m-i}{p \left(1-\frac{q}{p} \right)}$, which implies part (a) of the Lemma. On the other hand, setting $i=1$, if $\frac{q}{p}>1$ and $m$ is large enough, then $\mu_1$ is dominated by the term $\left( 1+\frac{1}{q-p}\right) \frac{\left( \frac{q}{p}\right)^{m} - \frac{q}{p}}{\frac{q}{p} - 1}$, which implies part (b) of the Lemma. \qed

\section{At least 4 types are needed for majority} \label{Section:Impossibility}

We begin by defining the \emph{rank} of a family of population protocols.

\begin{definition}[rank]
For any population protocol $P$, denote by $Q(P)$ the set of types used by $P$. 
Let ${\cal P}_{g}$ be a class of population protocols that stably compute the function $g$. 
The \emph{rank} of ${\cal P}_{g}$ is

\begin{equation}
R({\cal P}_{g}) \stackrel{def}{=} \min_{P \in {\cal P}_{g}} |Q(P)|.
\end{equation}  
\end{definition}

In this section, we prove that 3 states are not sufficient to stably compute the majority function in any 2-type population of agents and for any interaction graph.

\begin{theorem}
Let ${\cal P}_{majority}$ be the class of population protocols that stably compute the majority function in any 2-type population of agents and for any interaction graph. Then $R({\cal P}_{majority}) > 3$.
\end{theorem}
\proof Assume for the sake of contradiction that there is a population protocol $P \in {\cal P}_{majority}$ that uses only 3 types, namely $Q(P) = \{r, b, g\}$, among which $r$ and $g$ are input symbols. Notice also, that since we initially have a 2-type population, the final answer given by $P$ can have only two possible values (i.e.~indicating the type of the initial majority). Therefore, if the output function of $P$ assigns different values to each distinct state in $Q(P)$, then one of the states must never occur in a final configuration. Thus, we can set the output value of this state arbitrarily equal to the output value of one of the other two states. Consequently,we may assume that the output function of $P$ is $\gamma: Q(P) \to \{0, 1\}$. 

Since, by assumption $P \in {\cal P}_{majority}$, we cannot have $\gamma(r) = \gamma(b) = \gamma(g)$, otherwise $P$ would not be able to tell the difference between any two different input populations. Therefore, we may assume without loss of generality that there is a state $q_1 \in Q(P)$ such that $\gamma(q_1) = 1$, while for both the other two states $q_2$ and $q_3$ in $Q(P) \backslash \{q_1\}$, we have $\gamma(q_2) = \gamma(q_3) = 0$ (the case where $q_1$ is the only state having $\gamma(q_1) = 0$, while $\gamma(q_2) = \gamma(q_3) = 1$ is treated similarly, by symmetry). We can also assume without loss of generality, that when the initial majority is of type $r$, the protocol eventually reaches a configuration where all agents are of type $q_1$, and thus $P$ gives output 1. 

Assume now that we have a population $V$ of agents, among which $R \subseteq V$ are initially of type $r$ and $G \subseteq V$ are initially of type $g$. We will denote this configuration by $(R, G)$. By the above discussion, if $|V| = 2k+1$, for some integer $k \geq 2$, then $P$ will output 1 in both the following two configurations: (a) $C_1 = (R_1, G_1)$, where $|R_1| = k+1, |G_1| = k$ and (b) $C_2 = (R_2, G_2)$, where $|R_2| = k+2, |G_1| = k-1$. In particular, running $P$ on input either $C_1$ or $C_2$, we will eventually reach a configuration where all vertices are of type $q_1$ (in fact we will \emph{never} leave this configuration once we have reached it). In other words, there is a sequence of transitions $T_1$ (equivalently, there is a sequence of pairs of agents picked by the scheduler) that transforms configuration $C_1$ to the configuration where all agents are of type $q_1$. Similarly, there is a sequence of transitions $T_2$ that transforms configuration $C_2$ to the configuration where all agents are of type $q_1$.

Suppose now that we have a population $V' = V \cup \{v, u\}$, i.e.~$V'$ consists of $V$ together with two extra agents $v$ and $u$ (where $u, v \notin V$). Consider also the following two initial configurations: (a) $C'_1 = (R_1, G_1 \cup \{v, u\})$ and (b) $C'_2 = (R_2, G_2 \cup \{v, u\})$. In particular, in configuration $C'_1$, agents in $R_1$ are of type $r$ and agents in $G_1 \cup \{v, u\}$ are of type $g$. Furthermore, in configuration $C'_2$, agents in $R_2$ are of type $r$ and agents in $G_2 \cup \{v, u\}$ are of type $g$. Note that the majority in $C'_1$ is of type $g$, while the majority in $C'_2$ is of type $r$. Therefore, since $P \in {\cal P}_{majority}$, it follows that $P$ must output 0 when the starting configuration is $C'_1$ and 1 when the starting configuration is $C'_2$. But starting at $C'_1$ it is possible to follow the sequence of transitions $T_1$ (i.e.~ignoring agents $v$ and $u$ during these steps), thus reaching a configuration $C'$ where all agents in $V' \backslash \{v, u\}$ are of type $q_1$ and agents $v$ and $u$ remain of type $r$. Similarly, starting at $C'_2$ it is possible to follow the sequence of transitions $T_2$, thus reaching to the same configuration $C'$. This is a contradiction, since $P$ will not be able to tell the difference between the starting configurations $C'_1$ and $C'_2$. In particular, the output of $P$ after reaching $C'$ will be wrong for exactly one of the two initial configurations $C'_1$ or $C'_2$, contradicting the assumption that $P \in {\cal P}_{majority}$. \qed

\section{Computing majority in arbitrary interaction graphs}

In this section we introduce a symmetric population protocol with $4$ states (called the \emph{ambassador protocol}) which, given an \emph{arbitrary} undirected graph $G=(V,E)$ as the underlying interaction graph of the population, \emph{stably computes} the majority of the types of the vertices of $G$ (even if the majority differs only by one from the minority). Assuming that the input symbols are $g$ (for green) and $r$ (for red), the set of states in the ambassador protocol is $Q=\{(g,0),(g,1),(r,0),(r,1)\}$. The input function $\iota $ is such that $\iota(g)=(g,1)$ and $\iota(r)=(r,1)$. The output function $\gamma$ is such that $\gamma((g,i))=g$ and $\gamma((r,i))=r$, where $i \in \{0,1\}$. Finally, for simplicity of the presentation, we present the transition function $\delta$ in the form of a table in Figure~\ref{embassador-transition}.

\begin{figure}[tbh]
\centering 
$%
\begin{array}{|c|c|c|c|c|}
\hline
u\text{ }\backslash \text{ }v & (g,0) & (g,1) & (r,0) & (r,1) \\ \hline
\ \ (g,0) \ \  & - & \ \ \left( (g,1),(g,0)\right) \ \  & - & \ \ \left((r,1),(r,0)\right) \ \  \\ \hline
(g,1) & \ \ \left( (g,0),(g,1)\right) \ \  & - & \ \ \left((g,0),(g,1)\right) \ \  & \left((g,0),(r,0)\right) \\ \hline
(r,0) & - & \left( (g,1),(g,0)\right) & - & \left( (r,1),(r,0)\right) \\ \hline
(r,1) & \left( (r,0),(r,1)\right) & \left( (r,0),(g,0)\right) & \left((r,0),(r,1)\right) & - \\ \hline
\end{array}%
$%
\caption{The transition matrix of the ambassador model.}
\label{embassador-transition}
\end{figure}

The transition matrix of Figure~\ref{embassador-transition} can be interpreted as follows. Every row and every column is labeled by a state of $Q$. The cell that belongs to the row labeled with state $q_{1} \in Q$ and to the column labeled with state $q_{2}\in Q$ has either the symbol ``$-$'' or an ordered pair of states $(q'_1, q'_2) \in Q \times Q$. In the cases where this cell has the ordered pair $(q'_1, q'_2)$, then $\delta (q_1,q_2)=(q'_1, q'_2)$, otherwise $\delta (q_1, q_2) = (q_1, q_2)$. Note that the ambassador protocol is defined on undirected interaction graphs, i.e.~~in every interaction there is no initiator or responder. That is, whenever $\delta (q_1, q_2)=(q'_1, q'_2)$ then $\delta (q_2, q_1)=(q'_2, q'_1)$.

The main idea of the ambassador protocol can be described as follows. The first component of the state of a vertex $u$ denotes the color of $u$. That is, whenever $u$ is at state $(q,i)$ (resp. $(r,i)$), where $i\in \{0,1\}$, this is interpreted as ``$u$ is currently colored green (resp. red)''. Furthermore, if the second component of the state of $u$ is $1$ (resp. $0$), this is interpreted as ``$u$ has an ambassador'' (resp. ``$u$ has no ambassador''). Whenever two vertices $u$ and $v$ interact during the execution of the protocol, a peculiar battle takes place between the colors of $u$ and $v$, where the ``ambassadors'' of $u$ and $v$ (whenever they
exist) try to spread their own color to the other vertex, in the following sense:

\begin{itemize}
\item Assume that $u$ and $v$ have different colors. If both $u$ and $v$ have an ambassador, then they both keep their own colors in the next step, but both their ambassadors disappear (they die during their battle as they are equally ``strong''). If $u$ has an ambassador and $v$ has no ambassador, then $v$ takes the color of $u$ in the next step and the ambassador now moves from $u$ to $v$ (the ambassador wins the battle, as there is no opponent, and leaves vertex $u$ to conquer vertex $v$). If both $u$ and $v$ have no ambassador then their state remains the same at the next step (in this case no battle can take place).

\item Assume that both $u$ and $v$ have the same color; then they both maintain their color in the next step. If they both have (or if they both do not have) an ambassador, then their state remains the same at the next step (there is no battle whenever there is no ambassador, or between ambassadors of the same color). If one of them (say $u$) has an ambassador and the other one (say $v$) does not have one, then the ambassador moves from $u$ to $v$.
\end{itemize}

The correctness of the ambassador protocol under the assumption of an \emph{arbitrary} fair scheduler is proved in the next theorem.

\begin{theorem} \label{ambassador-fair-thm}
Given any (un)directed graph, if there exists initially a majority, then the $4$-state ambassador protocol stably computes the initial majority value.
\end{theorem}
\proof One of the crucial invariants of the protocol is that, at any step during its execution the difference between the number of ambassadors of the majority and the minority equals the difference between the initial majority and the initial minority. Indeed, whenever ambassadors disappear, they disappear in pairs, i.e.~one from each color.

For any $k, \ell \geq 0$ denote by $\mathcal{C}_{k, \ell}$ the set of configurations of $G$ with $k$ red ambassadors and $\ell $ green
ambassadors. Using the assumption that the scheduler is fair, we will prove
that, starting at a configuration $C\in \mathcal{C}_{k,\ell }$, where $k,\ell \geq 1$%
, $G$ will be led by the ambassador protocol in finite steps to a
configuration $C^{\prime }\in \mathcal{C}_{k-1,\ell -1}$. Assume otherwise
that, starting at such a configuration $C\in \mathcal{C}_{k,\ell }$, we stay
for ever at configurations in the set $\mathcal{C}_{k,\ell }$. First recall
that, whenever a vertex $u$ with a red ambassador interacts with a vertex $v$
that has a green ambassador, both the number of red and green ambassadors
decrease by one (cf. Figure \ref{embassador-transition}). Furthermore recall
that, whenever a vertex $u$ with an ambassador (of any color) interacts with
a vertex $v$ that has no ambassador, the ambassador of $u$ is moved to
vertex $v$ (and $v$ receives the color of $u$). Therefore, since $G$ is
connected and $k,\ell \geq 1$ by assumption, for every configuration in $%
\mathcal{C}_{k,\ell }$ there exists a chain of transitions that lead to a
configuration in $\mathcal{C}_{k-1,\ell -1}$. That is, every configuration
in $\mathcal{C}_{k,\ell }$ \emph{can lead} to a configuration in $\mathcal{C}%
_{k-1,\ell -1}$. Therefore, since the scheduler is fair, this will
eventually happen (in a finite number of steps).

Assume that the interaction graph $G$ has $n$ vertices and that the initial
majority has cardinality $k$ and the initial minority has cardinality $\ell $%
, where $k>\ell \geq 1$. Then, as we proved above, $G$ will be led to a
configuration in $\mathcal{C}_{k-i,\ell -i}$, for every $i=1,2,\ldots ,\ell $%
. That is, $G$ will be led in finite steps to a configuration in $\mathcal{C}%
_{k-\ell ,0}$, i.e.~to a configuration where all $k-\ell \geq 1$ ambassadors
are of the color of the initial majority. Note now that, once $G$ has a
configuration in $\mathcal{C}_{k-\ell ,0}$, it will stay for ever at a
configuration in $\mathcal{C}_{k-\ell ,0}$, since no ambassador can
disappear any more. Furthermore note by the definition of the ambassador
protocol (cf. Figure~\ref{embassador-transition}) that for every
configuration in $\mathcal{C}_{k,\ell }$ there exists a chain of transitions
that lead to a configuration where all vertices are colored with the color
of the initial majority (regardless of whether they have an ambassador or
not). Therefore, since the scheduler is fair, $G$ will reach eventually (in
a finite number of steps) a configuration where all vertices have the color
of the initial majority. Once $G$ has reached such a configuration, the
output function $\gamma $ will correctly compute the initial majority value
on every vertex. \qed

\medskip

The next theorem provides an upper bound on the expected running time of the protocol, until it stabilizes to the correct majority under the assumption of a probabilistic scheduler.

\begin{theorem} \label{ambassador-probabilistic-thm}
Let $G$ be an arbitrary connected undirected interaction graph with $n$ vertices. Assuming the probabilistic scheduler, if initially there are $k$ red vertices and $\ell \neq k$ green vertices, then the expected time until the $4$-state ambassador protocol converges is $O(n^6)$. If, additionally, the interaction graph is the complete graph $K_n$, then the expected time until the $4$-state ambassador protocol converges is $O\left(\frac{\ln{n}}{|k-\ell|} n^2 \right)$.
\end{theorem}
\proof Assume without loss of generality that $k <\ell$, i.e.~the green vertices are the initial majority. Let $T_1$ be the time needed for all red ambassadors to meet a green ambassador and thus disappear. Let also $T_2$ be the time after no red ambassadors remain in the graph, until the protocol stabilizes to the correct majority. Clearly, the time needed for the protocol to converge is $T_1+T_2$. 

We first consider an arbitrary undirected interaction graph $G$. In order to upper bound $\mathbb{E}[T_1]$, observe that, by the definition of our protocol and assuming the probabilistic scheduler, each red and green ambassador performs an independent random walk until it meets with an ambassador of the opposite color. By Corollary 1 of~\cite{TW91}, the expectation of the maximum meeting time is $O(n^3)$. If we also count the steps which do not cause a change in the configuration, we then have that the expected number of steps before two ambassadors of a different color meet (as long as there is at least one ambassador of each color) is at most $O(|E| n^3)$, where we used the fact that the expected number of steps between two steps that change the configuration is $O(|E|)$. Since we will have exactly $k$ meetings before all red ambassadors disappear, we then have that $\mathbb{E}[T_1] = O(k|E|n^3)$. Furthermore, in order to bound $\mathbb{E}[T_2]$, we can use Theorem 3 of~\cite{TW91}, which states that the expected number of steps needed for a single random walk to cover all vertices of a graph is $O(n^4)$. Combining this with the fact that the expected number of steps between two steps that change the configuration is $O(|E|)$, we have $\mathbb{E}[T_2] = O(|E|n^4)$, which implies the first part of the Theorem.

We now consider the case where $G = K_n$. In order to upper bound $\mathbb{E}[T_1]$, let $X_i, i= 1, \ldots, \ell-k$, be the time needed for the $i$-th pair of differently colored ambassadors to meet. Then $X_i$ is a geometric random variable with success probability $\frac{(\ell-i+1) (k-i+1)}{{n \choose 2}}$. Therefore, the expected time needed for all red ambassadors to disappear is 

\begin{eqnarray}
\mathbb{E}[T_1] & = & \sum_{i=1}^{k} \mathbb{E}[X_i] = {n \choose 2} \sum_{i=1}^{k} \frac{1}{(\ell-i+1) (k-i+1)} \\
& = & {n \choose 2} \frac{1}{\ell-k} \sum_{i=1}^{k} \left( \frac{1}{k-i+1} - \frac{1}{\ell-i+1} \right) \\
& \leq & {n \choose 2} \frac{1}{\ell-k} \sum_{i=1}^{k} \frac{1}{i} \\
& \leq & {n \choose 2} \frac{1+\ln{k}}{|k-\ell|}. 
\end{eqnarray}
In order to upper bound $\mathbb{E}[T_2]$, first note that, in the worst case, after all red ambassadors have disappeared, there will be $\ell-k$ green ambassadors and $n-(\ell-k)$ red vertices with no ambassador. Let $Y_i, i=1, \ldots, n-(\ell-k)$ be the time needed for the $i$-th red vertex to become green by interacting with a green ambassador. Then $Y_i$ is a geometric random variable with success probability $\frac{(\ell-k) (n-(\ell-k) -i+1)}{{n \choose 2}}$. Therefore, we have that

\begin{eqnarray}
\mathbb{E}[T_2] & = & \sum_{i=1}^{n-(\ell-k)} \mathbb{E}[Y_i] = {n \choose 2} \sum_{i=1}^{n-(\ell-k)} \frac{1}{(\ell-k) (n-(\ell-k) -i+1)} \\
& = & {n \choose 2} \frac{1}{\ell-k} \sum_{i=1}^{n-(\ell-k)} \frac{1}{i} \\
& \leq & {n \choose 2} \frac{1+\ln{n}}{|k-\ell|}. 
\end{eqnarray}
This completes the proof. \qed

\medskip

From the above Theorem, we can see that the running time of our 4-state protocol in the case where the difference between the majority and the minority is $\Theta(n)$ is $O(n \ln{n})$, which is comparable to the running time of the fast 3-state protocol of~\cite{AAE08}.

\section{The model of Angluin et al. in arbitrary interaction graphs}

In this section, we provide a detailed analysis of the 3-state protocol of~\cite{AAE08} on arbitrary interaction graphs $G$. In particular, in Subsection \ref{Section:Random}, we present our result concerning the random initial placement of individuals on the vertices of the interaction graph. In Subsection \ref{Section:Clique}, we prove our auxiliary result that, when the minority is sufficiently small, the probability that the protocol of Angluin et al. fails in computing the majority value is exponentially small. Although this result shows the robustness of the protocol of~\cite{AAE08} in the clique, we use it as an intermediate step in proving in Subsection \ref{Section:Minority domination} that there exists a family of graphs in which the protocol can fail with high probability. Finally, in Subsection \ref{Section:Exponential}, we prove the existence of a family of graphs in which the protocol of~\cite{AAE08} can take an exponential expected number of steps to reach consensus.

\subsection{Random initial placement} \label{Section:Random}

We prove in this subsection a preliminary result concerning the model of~\cite{AAE08} when the interaction graph is an arbitrary, strongly connected, directed graph $G$. In particular, we prove that if the initial assignment of individuals to the vertices of $G$ is random, then the majority protocol described in~\cite{AAE08} correctly identifies the initial majority with probability at least $\frac{1}{2}$. 

For the proof of Theorem \ref{highermajority} below, we will need a result from extremal combinatorics concerning systems of distinct representatives. A \emph{system of distinct representatives} for a sequence of (not necessarily distinct) sets ${\cal T}_1, {\cal T}_2, \ldots, {\cal T}_x$ is a sequence of distinct elements $e_1, e_2, \ldots, e_x$ such that $e_i \in {\cal T}_i$ for all $i = 1, 2, \ldots, x$. The following is a consequence of Hall's marriage Theorem (for a proof see chapter 5 in~\cite{Jbook}):

\begin{corollary}[\hspace{-0.01cm}\cite{Jbook}] \label{CorJbook}
Let ${\cal T}_1, {\cal T}_2, \ldots, {\cal T}_x$ be $r$-element subsets of a universe of $y$ elements, such that each element belongs to the same number $d \geq 1$ of these sets. If $x \leq y$, then the sets ${\cal T}_1, {\cal T}_2, \ldots, {\cal T}_x$ have a system of distinct representatives.
\end{corollary}

We are now ready to prove the following Theorem.

\begin{theorem} \label{highermajority}
For any strongly connected directed graph $G$, if the initial assignment of individuals to the vertices of $G$ is random, then the majority protocol described in~\cite{AAE08} correctly identifies the initial majority with probability at least $\frac{1}{2}$.
\end{theorem}
\proof Assume without loss of generality that the initial majority is of type $g$ (green), i.e.~$|G_0| \geq |R_0|$. The state space of the Markov chain ${\cal M}$ describing the evolution of the protocol will be the set ${\cal S} = \{(A, B): A, B \subseteq V, A\cap B = \emptyset\}$. In particular, when ${\cal M}$ is in state $(A, B)$ at time $t$, we will write that $(R_t, G_t) = (A, B)$, and this will mean that the subset of vertices $A$ (respectively $B$) is of type $r$ (respectively $g$). It is also evident that the set of vertices $V-(A\cup B)$ is of type $b$. It then follows that the majority protocol in~\cite{AAE08} fails to correctly identify type $g$ as the initial majority if ${\cal M}$ eventually reaches state $(V, \emptyset)$; we will then say that the \emph{initial minority wins} (whereas in the opposite case we will say that the \emph{initial majority wins}). Since the initial placement of individuals on the vertices of the graph is random and the initial minority is of type $r$, we also have that

\begin{equation} \label{minoritysum}
\Pr\{\textrm{initial minority wins}\} = \sum_{S \subseteq V: |S| = |G_0|} \frac{1}{{n \choose |G_0|}} \Pr\{\textrm{$(V, \emptyset)$ reached} | (R_0, G_0) = (V-S, S)\}.  
\end{equation}
Consider now a fixed choice of $S$ in the above sum and an arbitrary set $A \subseteq S$ of size $|A| = |R_0| = |V - S|$ (notice that it is always possible to find such a set $A$, since $|S| \geq |V-S|$, by assumption). Then we have 

\begin{eqnarray}
\Pr\{\textrm{$(V, \emptyset)$ reached} | (R_0, G_0) = (V-S, S) \} & \leq & \Pr\{\textrm{$(V, \emptyset)$ reached} | (R_0, G_0) = (V-A, A) \} \\
& = & \Pr\{\textrm{$(\emptyset, V)$ reached} | (R_0, G_0) = (A, V-A) \}. \label{dominationeq}
\end{eqnarray}
For the inequality we used the fact that $V-S \subseteq V-A$ and that increasing the initial number of agents of type $r$ increases the probability that agents of type $r$ win. For the equality we used the fact that the protocol is symmetric for types $g$ and $r$. In words, (\ref{dominationeq}) states that the probability that the initial minority wins, starting from $(V-S, S)$, is at most the probability that the initial majority wins if we exchange agents of type $r$ initially placed in $V-S$ with agents of type $g$ initially placed in $A$.

Notice now that if for any choice of $S$ we assign a unique choice of $A = A_S \subseteq S$, i.e.~such that for any $S \neq S'$ we have $A_S \neq A_{S'}$, then we are done, since by (\ref{minoritysum}) and (\ref{dominationeq}) we have

\begin{eqnarray}
\Pr\{\textrm{initial minority wins}\} & = & \sum_{S \subseteq V: |S| = |G_0|} \frac{1}{{n \choose |G_0|}} \Pr\{\textrm{$(V, \emptyset)$ reached} | (R_0, G_0) = (V-S, S)\} \\
& \leq & \sum_{S \subseteq V: |S| = |G_0|} \frac{1}{{n \choose |G_0|}} \Pr\{\textrm{$(\emptyset, V)$ reached} | (R_0, G_0) = (A_S, V-A_S)\} \\
& \leq & \Pr\{\textrm{initial majority wins}\}.
\end{eqnarray}

We now show that such a one to one correspondence of sets $A_S$ to sets $S$ is possible: For $k = |R_0|$, let $T_1, T_2, \ldots, T_{n \choose k}$ be an arbitrary enumeration of subsets of $V$ of size exactly $k$. For any $1 \leq i \leq {n \choose k}$, let also ${\cal T}_i$ be the set of subsets of $V-T_i$ of size $k$. Notice that the set ${\cal T}_i$ is non-empty (because $n-k \geq k$ by assumption) and is also uniquely determined by $T_i$. Furthermore, the sets ${\cal T}_i$ have the same size $r = {n-k \choose k}$ and each set $T_j$ belongs to exactly $d = {n-k \choose k}$ distinct ${\cal T}_i$'s. Therefore, we can apply Corollary \ref{CorJbook} to the sets ${\cal T}_i$, with $x = y = {n \choose k}$. In particular, this implies that the sets ${\cal T}_i$, $1 \leq i \leq {n \choose k}$, have a system of distinct representatives, which concludes the proof. \qed

\medskip

In the above theorem we provided a sufficient condition under which the majority protocol described in~\cite{AAE08} correctly identifies the initial majority with probability at least $\frac{1}{2}$. This result is in wide contrast to the negative result of Subsection~\ref{Section:Minority domination} (cf.~Theorem~\ref{theorem-minority-domination}), in which we highlight a case where the majority protocol of~\cite{AAE08} fails with high probability.

\subsection{Clique} \label{Section:Clique}

In this Subsection we provide an upper bound on the probability that all agents eventually become of type $r$, given that we start with $\epsilon n$ agents of type $r$ and $(1-\epsilon) n$ agents of type $g$ on the $n$-vertex clique, where $\frac{1}{n} \leq \epsilon < \frac{1}{7}$ (the upper bound on $\epsilon$ is used to facilitate the exposition of our arguments). We assume without loss of generality that $\epsilon n$ is an integer. By the discussion in Subsection \ref{representation}, the state space of the Markov chain ${\cal M}$ describing the evolution of the protocol at time $t$ contains tuples of the form $(|R_t|, |G_t|)$, where $R_t$ (resp. $G_t$) is the set of vertices of type $r$ (resp. $g$) at time $t$. In particular, we are interested in upper bounding 

\begin{equation}
\Pr\{\textrm{absorption at $(n, 0)$} | \textrm{initially at $(\epsilon n, n-\epsilon n)$}\}.
\end{equation}

The core of our proof lies in the definition of two discrete time processes ${\cal W}$ and ${\cal C}$ that ``filter'' the information from the original Markov chain ${\cal M}$. 

\begin{definition}[The Blank Process ${\cal W}$]
This process keeps track of the number of blank vertices over time, i.e.~${\cal W}(t) \stackrel{def}{=} \left\langle \textrm{$\#$ vertices of type $b$ at time $t$}\right\rangle$.
\end{definition}

For convenience, we will use the following notation to describe transitions of ${\cal M}$: We will write $g \to r$ to describe a transition of the form $(x, y) \to (x-1, y)$, for some $x, y \in \{1, 2, \ldots, n\}$. More specifically, $g \to r$ is used to describe a transition where a directed edge $(v, u)$ is chosen by the scheduler, $v$ is of type $g$, and $u$ is of type $r$. Similarly, we will use $r \to g$ for transitions of the form $(x, y) \to (x, y-1)$, $g\to b$ for transitions of the form $(x, y) \to (x, y+1)$ and $r\to b$ for transitions of the form $(x, y) \to (x+1, y)$. We note that the state of ${\cal M}$ at any time $t$ can be fully described by the initial state and by a sequence of transitions among $\{g \to r, r \to g, g \to b, r \to b\}$.

\begin{definition}[The Contest Process ${\cal C}$]
Transitions of ${\cal M}$ are paired recursively, starting from time 0 as follows: Every transition that increases the number of blanks is paired with the earliest subsequent transition that decreases the number of blanks and is not paired yet. \footnote{We assume that the pairing concerns only transitions that change the state of ${\cal M}$. In particular, transitions of the form $b \to r, b \to g, b \to b, g \to g$ and $r \to r$ are ignored in this pairing as irrelevant.} For an arbitrary time $t$, we denote by $\tau(t)$ (or just $\tau$ for short) the number of pairs until time $t$. The Contest Process ${\cal C}$ is defined over the time scale $\tau$, where ${\cal C}(0) = |R_0|$ and for $\tau = 1, 2, \ldots,$

\begin{equation}
{\cal C}(\tau) = \left\{ 
\begin{array}{ll}
	{\cal C}(\tau-1) + 1, & \quad \textrm{if the $\tau$-th pair is $(r \to g, r \to b)$,} \\
	{\cal C}(\tau-1) - 1, & \quad \textrm{if the $\tau$-th pair is $(g\to r, g\to b)$ and} \\
	{\cal C}(\tau-1), & \quad \textrm{otherwise.}
\end{array}
\right.
\end{equation} 
\end{definition}

For example, suppose that the underlying clique has $n=4$ vertices and we have the sequence of states $S_0 = (1, 3), S_1 = (1, 2), S_2 = (1, 3), S_3 = (1, 2), S_4 = (1, 1), S_5 = (2, 1), S_6 = (2, 0), S_7 = (3, 0), S_8 = (4, 0)$. In the new notation, we start at $S_0 = (1, 3)$ and then we have the transitions $(1)~r \to g, (2)~g \to b, (3)~r \to g, (4)~r \to g, (5)~r \to b, (6)~r \to g, (7)~r \to b, (8)~r \to b$. The value of ${\cal W}_t$, for any $t = 0, \ldots, 8$ is then easy to find (e.g. ${\cal W}_5 = 1$). For the Contest process ${\cal C}$, we pair the transitions as follows: (a) $S_0 \to S_1$ (which is $r \to g$) is paired to $S_1 \to S_2$ (which is $g \to b$), (b) $S_2 \to S_3$ (which is $r \to g$) is paired to $S_4 \to S_5$ (which is $r \to b$), (c) $S_3 \to S_4$ (which is $r \to g$) is paired to $S_6 \to S_7$ (which is $r \to b$) and (d) $S_5 \to S_6$ (which is $r \to g$) is paired to $S_7 \to S_8$ (which is $r \to b$). In particular, there are only 4 transitions for ${\cal C}$ and in particular, we have ${\cal C}(0) = 1, {\cal C}(1) = 1, {\cal C}(2) = 2, {\cal C}(3) = 3, {\cal C}(4) = 4$.

Notice that the processes ${\cal W}$ and ${\cal C}$ are dependent. As a matter of fact, ${\cal C}$ is not even defined using the same time scale as ${\cal W}$ and ${\cal M}$ (to indicate this, we have used the convention that $t$ is the time variable for processes ${\cal M}, {\cal W}$, while $\tau$ is the time variable for process ${\cal C}$). However, observe that if we initially begin with no blanks (i.e.~$|R_0|+|G_0|=n$ hence ${\cal W}(0) = 0$), then whenever ${\cal W}$ decreases its value, we have a transition step of ${\cal C}$. Additionally, we can prove the following:

\begin{lemma}[Relating ${\cal C}$ and ${\cal M}$] \label{relation-lemma}
For any $T \in \mathbb{N}$, denote by ${\cal C}_{|T}$ the value of ${\cal C}$ given only states ${\cal M}(t), t=0, 1, \ldots, T$ (i.e.~given the history of ${\cal M}$ up to time $T$). Then, ${\cal C}_{|T} \geq |R_T|$ for any $T \in \mathbb{N}$. Furthermore, if ${\cal C}_{|T} = 0$, then all vertices are of type $g$.
\end{lemma}
\proof Notice that if \emph{all} transitions up to time $T$ were paired according to the pairing in the definition of ${\cal C}$, then we would have exactly ${\cal C}_{|T} = |R_T|$. Indeed, each pair of transitions of the form $(r \to g, r \to b)$ increases the number of vertices of type $r$ by 1, each pair of transitions of the form $(g \to r, g \to b)$ decreases the number of vertices of type $r$ by $1$ and any pair of transitions of the form $(r \to g, g \to b)$ or $(g \to r, r \to b)$ does not change the number of vertices of type $r$. Finally, notice that transitions that are not paired are either of the form $r \to g$, or $g \to r$, which can only decrease the number of vertices of type $r$. This completes the proof of the first part.

For the second part of the Lemma, let $T_0$ be a time where ${\cal C}_{|T_0} = 0$ and note that because of the first part, we already have that $|R_t| = 0$, so we only need to prove that also ${\cal W}(T_0) = 0$. Assume for the sake of contradiction that there is some vertex $w$ that is of type $b$ at time $T_0$. By definition, this implies that there is an unpaired transition of the form $r \to g$ or $g \to r$. Indeed, this follows from the observation that paired transitions do not change the number of vertices of type $b$, therefore if all transitions are paired there must be no vertices of type $b$ remaining. If now the unpaired transition is of the form $r \to g$, then the existence of a vertex of type $r$ at the time of the transition (say time $t < T_0$) together with the fact that $|R_{T_0}|=0$ would imply that there is also an unpaired subsequent transition (i.e.~a transition that happened at some time $t'$, with $T_0 > t' >  t$) of the form $g \to r$. But if there is an unpaired transition of the form $g \to r$, then pairing it with a subsequent transition of the form $g \to b$ after time $T_0$ would decrease the value of ${\cal C}'$ by 1, making it negative, which leads to a contradiction because of the first part of the Lemma. This completes the proof of the second part. \qed

\medskip

We will use the following domination statements in Lemmas \ref{Blank-domination-lemma} and \ref{Contest-domination-lemma}, which concern the domination of processes ${\cal W}$ and ${\cal C}$ by appropriate birth-death processes:

\begin{lemma}[Domination of ${\cal W}$] \label{Blank-domination-lemma}
Let $\alpha, \beta, \kappa \in \{1, \ldots, n-1\}$, with $\alpha< \beta$. Let also ${\cal B}_{\cal W}$ be a birth-death process, which has state space ${\cal S_{{\cal B}_{\cal W}}} = \{S_0, \ldots, S_n\}$, with $S_n$ an absorbing state and transition probability matrix $P$, with $P(S_i, S_{i+1})=1$ for all $i \in \{0, \ldots, \alpha\} \cup \{\beta, \ldots, n-1\}$, $\frac{P(S_i, S_{i+1})}{P(S_i, S_{i-1})} = \frac{2\kappa}{\alpha}$ for all $i \in \{\alpha+1, \ldots, \beta-1\}$ and $P(S_i, S_i) = \Pr({\cal W}(t) = i| {\cal W}(t-1) = i)$, for all $t \geq 1$ and for all $i \in \{\alpha+1, \ldots, \beta-1\}$. Then, given that the vertices of type $r$ are at most $\kappa$, the process ${\cal W}$ is stochastically dominated by ${\cal B}_{\cal W}$ in the following sense: $\Pr({\cal W}(t) > x| {\cal W}(0)=0) \leq \Pr({\cal B}_{\cal W}(t) \in \cup_{y>x} S_y| {\cal B}_{\cal W}(0)=S_0)$, for any time $t$ and $x \in \{0, \ldots, n\}$.
\end{lemma}
\proof For the proof, it suffices to show that, for any $t \geq 1$ and for any $i \in \{0, \ldots, n-1\}$, 

\begin{equation}
\Pr({\cal W}(t) = i+1| {\cal W}(t-1)=i) \leq \Pr({\cal B}_{\cal W}(t) = S_{i+1}| {\cal B}_{\cal W}(t-1) = S_i).
\end{equation}
This is trivially true for all $i \in \{0, \ldots, \alpha\} \cup \{\beta, \ldots, n-1\}$, because the right hand side of the above inequality is 1. For $i \in \{\alpha+1, \ldots, \beta-1\}$, given that ${\cal W}(t-1)=i$ (i.e.~there are exactly $i$ blanks at time $t$), the probability that ${\cal W}$ increases by 1 in the next time step is equal to the probability that either a transition $g \to r$ or a transition $r \to g$ occurs, which is equal to $\frac{2|R_{t-1}| |G_{t-1}|}{n(n-1)}$. On the other hand, given that ${\cal W}(t-1)=i$, the probability that ${\cal W}$ decreases by 1 in the next time step is equal to the probability that either a transition $g \to b$ or a transition $r \to b$ occurs, which is equal to $\frac{i(|R_{t-1}| + |G_{t-1}|)}{n(n-1)}$. But then 

\begin{equation}
\frac{\Pr({\cal W}(t) = i+1| {\cal W}(t-1)=i)}{\Pr({\cal W}(t) = i-1| {\cal W}(t-1)=i)} = \frac{2|R_{t-1}| |G_{t-1}|}{i(|R_{t-1}| + |G_{t-1}|)} \leq \frac{2 |R_{t-1}|}{i}.
\end{equation}
By assumption, this is at most $\frac{2 \kappa}{\alpha}$, which combined with the fact that $P(S_i, S_i) = \Pr({\cal W}(t) = i| {\cal W}(t-1) = i)$, for all $i \in \{\alpha+1, \ldots, \beta-1\}$, concludes the proof. \qed

\begin{lemma}[Domination of ${\cal C}$] \label{Contest-domination-lemma}
Let $\beta, \kappa$ be positive integers, with $\beta+\kappa<n$. Let also ${\cal B}_{\cal C}$ be a birth-death process, which has state space ${\cal S_{{\cal B}_{\cal C}}} = \{T_0, \ldots, T_n\}$, with $T_0, T_n$ absorbing states and transition probability matrix $Q$, with $Q(T_i, T_{i+1}) = 1$ for all $i \in \{\kappa, \ldots, n-1\}$, $\frac{Q(T_i, T_{i+1})}{Q(T_i, T_{i-1})} = \frac{\kappa}{n-\beta-\kappa}$ for all $i \in \{1, \ldots, \kappa-1\}$ and $Q(T_i, T_i) = \Pr({\cal C}(\tau) = i| {\cal C}(\tau-1) = i)$, for all $\tau \geq 1$ and for all $i \in \{1, \ldots, \kappa-1\}$. Then, given that the vertices of type $b$ are at most $\beta$, the process ${\cal C}$ is stochastically dominated by ${\cal B}_{\cal C}$ in the following sense: $\Pr({\cal C}(\tau) > x| {\cal C}(0)=|R_0|) \leq \Pr({\cal B}_{\cal C}(\tau) \in \cup_{y>x} T_y| {\cal B}_{\cal C}(0)=|R_0|)$, for any $\tau$ and $x \in \{0, \ldots, n\}$.
\end{lemma}
\proof It suffices to show that, for any $\tau \geq 1$ and for any $i \in \{1, \ldots, n-1\}$, 

\begin{equation}
\Pr({\cal C}(\tau) = i+1| {\cal C}(\tau-1)=i) \leq \Pr({\cal B}_{\cal C}(\tau) = T_{i+1}| {\cal B}_{\cal C}(\tau-1) = T_i).
\end{equation}
This is trivially true for all $i \in \{\kappa, \ldots, n-1\}$, because the right hand side of the above inequality is 1. For $i \in \{1, \ldots, \kappa-1\}$, we apply the \emph{principle of deferred decisions}. In particular, let $t_1$ (resp. $t_2$) be the time in the time scale of ${\cal M}$ that corresponds to the first (resp. second) transition of the $\tau$-th transition pair in the definition of ${\cal C}$ (notice that both $t_1$ and $t_2$ are random variables). Given that ${\cal C}(\tau-1)=i$, the probability that ${\cal C}$ increases by 1 in the next time step in the time scale of ${\cal C}$ is equal to the probability that the $\tau$-th transition pair is $(r \to g, r \to b)$, which is equal to $\Pr({\cal C}(\tau) = i+1| {\cal C}(\tau-1)=i, t_1, t_2) = \frac{|R_{t_1}||G_{t_1}|}{2 |R_{t_1}||G_{t_1}|} \frac{|R_{t_2}| (n-|R_{t_2}|-|G_{t_2}|)}{(|R_{t_2}|+|G_{t_2}|) (n-|R_{t_2}|-|G_{t_2}|)}$. On the other hand, the probability that the $\tau$-th transition pair in the definition of ${\cal C}$ is $(g \to r, g \to b)$ is $\Pr({\cal C}(\tau) = i-1| {\cal C}(\tau-1)=i, t_1, t_2) = \frac{|G_{t_1}||R_{t_1}|}{2 |R_{t_1}||G_{t_1}|} \frac{|G_{t_2}| (n-|R_{t_2}|-|G_{t_2}|)}{(|R_{t_2}|+|G_{t_2}|) (n-|R_{t_2}|-|G_{t_2}|)}$. But then 

\begin{equation}
\frac{\Pr({\cal C}(\tau) = i+1| {\cal C}(\tau-1)=i, t_1, t_2)}{\Pr({\cal C}(\tau) = i-1| {\cal C}(\tau-1)=i, t_1, t_2)} = \frac{|R_{t_2}|}{|G_{t_2}|}.
\end{equation}
Since $i \leq \kappa-1$, by assumption, it follows by Lemma \ref{relation-lemma} that $|R_{t_2}| < \kappa$. Therefore, we have that $\frac{\Pr({\cal C}(\tau) = i+1| {\cal C}(\tau-1)=i)}{\Pr({\cal C}(\tau) = i-1| {\cal C}(\tau-1)=i)} \leq \frac{\kappa}{n-\beta-\kappa}$, which combined with the fact that $Q(T_i, T_i) = \Pr({\cal C}(\tau) = i| {\cal C}(\tau-1) = i)$, for all $i \in \{1, \ldots, \kappa-1\}$, concludes the proof. \qed

\medskip

We are now ready to prove our main Theorem, which is stated below.

\begin{theorem} \label{theorem-clique}
Let $\epsilon < \frac{1}{7}$. For large enough $n$, starting from $\epsilon n$ agents of type $r$ and $(1-\epsilon) n$ agents of type $g$ on the clique $K_n$, the probability that the clique eventually contains only agents of type $r$ is at most $e^{-\Theta(n)}$.
\end{theorem}
\proof For the proof, we will upper bound the probability that the initial minority wins by providing upper bounds to the probability that the following events occur, in which $\kappa, \beta$ are predefined integers (which we fix later to facilitate exposition):

\begin{description}

\item[(i)] $A_1$ is the event that ${\cal C}$ reaches $\kappa$ before the number of vertices of type $b$ reaches $\beta$.

\item[(ii)] $A_2$ is the event that the number of vertices of type $b$ reaches $\beta$ before ${\cal C}$ reaches $\kappa$.

\end{description}
In particular, we note that 

\begin{equation} \label{eq-minority}
\Pr\{\textrm{initial minority wins}| {\cal M}(0) = (\epsilon n, n-\epsilon n)\} \leq \Pr(A_1 \cup A_2| {\cal M}(0) = (\epsilon n, n-\epsilon n)) 
\end{equation}
so we need to provide upper bounds for $\Pr(A_1| {\cal M}(0) = (\epsilon n, n-\epsilon n))$ and $\Pr(A_2| {\cal M}(0) = (\epsilon n, n-\epsilon n))$.
 
We now set $\alpha = \frac{n}{3}$, $\beta = \frac{n}{2}$ and $\kappa = \frac{n}{7}$. In that case, the forward bias of the birth-death process ${\cal B}_{\cal W}$ at states $\{S_{\alpha+1}, S_{\alpha+2} \ldots, S_{\beta-1}\}$ is at most $r_{\cal W} \stackrel{def}{=} \frac{6}{7} < 1$ and the forward bias of the birth-death process ${\cal B}_{\cal C}$ at states $\{T_1, T_2, \ldots, T_{\kappa-1}\}$ is at most $r_{\cal C} \stackrel{def}{=} \frac{2}{5} < 1$. 

In particular, by Lemma \ref{Contest-domination-lemma} and Lemma \ref{Birth-death-absorption}, given that the number of vertices of type $b$ is less than $\beta$, the probability that the process ${\cal C}$ reaches value $\kappa$ before reaching 0, given that it starts at $\epsilon n$, is at most $\frac{\left( \frac{1}{r_{\cal C}} \right)^{\epsilon n}-1}{\left(\frac{1}{r_{\cal C}}\right)^{\frac{n}{7}}-1}$. Therefore,  

\begin{equation} \label{eq-A1}
\Pr(A_1| {\cal M}(0) = (\epsilon n, n-\epsilon n)) \leq \frac{\left( \frac{1}{r_{\cal C}} \right)^{\epsilon n}-1}{\left(\frac{1}{r_{\cal C}}\right)^{\frac{n}{7}}-1} = \frac{\left(\frac{5}{2}\right)^{\epsilon n}-1}{\left(\frac{5}{2}\right)^{\frac{n}{7}}-1}.
\end{equation}

For the bound on $\Pr(A_2| {\cal M}(0) = (\epsilon n, n-\epsilon n))$, notice that, given ${\cal M}(0) = (\epsilon n, n-\epsilon n)$, the event $A_2$ can only happen if ${\cal W}$ reaches $\beta$ before ${\cal C}$ reaches either 0 or $\kappa$. For integers $k_0$ and $k_1$ we will define the following events: (i) $B_1$ is the event that ${\cal W}$ reaches $\beta$ and backtracks less than $k_0$ times until then and (ii) $B_2$ is the event that ${\cal W}$ reaches $\beta$ and ${\cal C}$ takes more than $k_1$ steps (in the time scale of ${\cal C}$) to reach either value 0 or $\kappa$. We will set $k_0 = k_1 = \left(\frac{1}{r_{\cal W}}\right)^{\frac{n}{12}} = \left(\frac{7}{6}\right)^{\frac{n}{12}}$ (however, note here that equation (\ref{eq-A2A1c}) remains valid for any $k_1 \leq k_0$). By observing that, every time ${\cal W}$ backtracks, we have a step in ${\cal C}$, we have that

\begin{equation} \label{eq-A2A1c}
\Pr(A_2| {\cal M}(0) = (\epsilon n, n-\epsilon n)) \leq \Pr(B_1 \cup B_2| {\cal M}(0) = (\epsilon n, n-\epsilon n)).
\end{equation} 
Indeed, given ${\cal M}(0) = (\epsilon n, n-\epsilon n)$, the probability that ${\cal W}$ reaches $\beta$ before ${\cal C}$ reaches either 0 or $\kappa$ is at most the probability of the event that ${\cal W}$ reaches $\beta$ and backtracks less than $k_0$ times until then (i.e.~$B_1$) or that ${\cal W}$ reaches $\beta$ and backtracks at least $k_0$ times but it takes longer for ${\cal C}$ to reach either 0 or $\kappa$ (i.e.~$B_2 \cap \overline{B_1}$).  

Notice now that, by Lemma \ref{Blank-domination-lemma} and Lemma \ref{Birth-death-absorption}, the probability that the process ${\cal W}$ reaches $\beta$ before going back to $\alpha$, given that it starts with $\alpha+1$ blank vertices is at most $\frac{\frac{1}{r_{\cal W}}-1}{\left(\frac{1}{r_{\cal W}}\right)^{\frac{n}{6}}-1}$. Therefore, by coupling ${\cal W}$ and ${\cal B}_{\cal W}$ (defined in Lemma \ref{Blank-domination-lemma}), we have that

\begin{eqnarray}
& & \Pr(B_1| {\cal M}(0) = (\epsilon n, n-\epsilon n)) \nonumber \\
& & \quad \leq \Pr\{\textrm{${\cal B}_{\cal W}$ reaches $S_{\beta}$ and backtracks less than $k_0$ times} | {\cal B}_{\cal W}(0) = \alpha+1\} \label{eq7} \\
& & \quad \leq \Pr\{\textrm{${\cal B}_{\cal W}$ reaches $S_{\beta}$ and visits $S_{\alpha}$ less than $k_0$ times} | {\cal B}_{\cal W}(0) = \alpha+1\} \label{eq8} \\
& & \quad \leq 1 - \left( 1 - \frac{\frac{1}{r_{\cal W}}-1}{\left(\frac{1}{r_{\cal W}}\right)^{\frac{n}{6}}-1} \right)^{k_0} \leq k_0 \left(\frac{1}{r_{\cal W}}\right)^{-\frac{n}{6}} = \left(\frac{7}{6}\right)^{-\frac{n}{12}}. \label{eq-B1}
\end{eqnarray}
Inequality (\ref{eq7}) follows by domination in Lemma \ref{Blank-domination-lemma}. The first inequality in (\ref{eq-B1}) follows from the memoryless property of Markov chains and the observation that every time ${\cal B}_{\cal W}$ takes the value $\alpha$ it immediately takes the value $\alpha+1$ on the next step, since it has a reflecting barrier on $S_{\alpha}$. For the second inequality in (\ref{eq-B1}) we used the fact that $(1-x)^y \geq 1-2xy$ whenever $0< xy<\frac{1}{2}$.

Let now $X_{\cal C}$ be the number of steps (in the time scale of ${\cal C}$) needed for process ${\cal C}$ to reach either value 0 or $\kappa$. Notice also that we can use Lemma \ref{Birth-death-Time}(a) in order to upper bound the expected time needed for ${\cal B}_{\cal C}$ to reach either $T_0$ or $T_{\kappa}$. In particular, this is at most the time needed for ${\cal B}_{\cal C}$ to reach $T_0$ if it had a reflecting barrier at $T_{\kappa}$, which is a mirror birth-death process to that of Lemma \ref{Birth-death-Time}(a) if we ignore all states other than $T_1, \ldots, T_{\kappa}$ (i.e.~with $m=\kappa$ and $S_i = T_{\kappa-i}, i =0, \ldots, \kappa$). Therefore, the expected time needed for ${\cal B}_{\cal C}$ to reach either $T_0$ or $T_{\kappa}$, given that it starts at $T_{\epsilon n}$ is $O\left( \frac{\epsilon n |E|^2}{\frac{1}{r_{\cal C}} -1}\right) = O(n^5)$, where we also used the fact that the probability that ${\cal C}$ (and thus also ${\cal B}_{\cal C}$) does not stay on the same state, given that it starts at one of the states $T_1, \ldots, T_{\kappa -1}$, is at least $\Omega\left(\frac{1}{|E|^2}\right)$ (in particular, this is a lower bound on the value of $p$ in Lemma \ref{Birth-death-Time}(a)). Therefore, by Lemma \ref{Contest-domination-lemma} and taking into consideration that ${\cal C}(0)= \epsilon n$, the expected value of $X_{\cal C}$ satisfies

\begin{equation} \label{eq10}
\mathbb{E}[X_{\cal C}| {\cal C}(0)=\epsilon n] = O(n^5).
\end{equation}
By Markov's inequality we then have, for large enough $n$,

\begin{equation} \label{eq11}
\Pr\left(\left. X_{\cal C} > k_1 \right| {\cal C}(0)=\epsilon n \right) < \frac{\mathbb{E}[X_{\cal C} | {\cal C}(0)=\epsilon n]}{k_1} = \frac{O(n^5)}{k_1}.
\end{equation}
In particular, we have that

\begin{equation} \label{eq-B2-B1}
\Pr(B_2 \cap \overline{B_1}| {\cal M}(0)=(\epsilon n, n-\epsilon n)) \leq \Pr\left(\left. X_{\cal C} > k_1 \right| {\cal C}(0)=\epsilon n \right) = \frac{O(n^5)}{k_1} \leq \left(\frac{7}{6}\right)^{-\frac{n}{13}}.
\end{equation}

By using now the bounds we have for $\Pr(A_1| {\cal M}(0) = (\epsilon n, n-\epsilon n))$ in (\ref{eq-A1}), $\Pr(B_1| {\cal M}(0) = (\epsilon n, n-\epsilon n))$ in (\ref{eq-B1}), and $\Pr(B_2 \cap \overline{B_1}| {\cal M}(0) = (\epsilon n, n-\epsilon n))$ in (\ref{eq-B2-B1}), together with inequalities (\ref{eq-minority}) and (\ref{eq-A2A1c}), we get that

\begin{eqnarray} 
\Pr\{\textrm{initial minority wins}| {\cal M}(0) = (\epsilon n, n-\epsilon n)\} & \leq & \frac{\left(\frac{1}{r_{\cal C}}\right)^{\epsilon n}-1}{\left(\frac{1}{r_{\cal C}}\right)^{\frac{n}{7}}-1} + k_0 \left(\frac{1}{r_{\cal W}}\right)^{-\frac{n}{6}} + \frac{O(n^5)}{k_1} \label{fin-eq} \\
& \leq & \frac{\left(\frac{5}{2}\right)^{\epsilon n}-1}{\left(\frac{5}{2}\right)^{\frac{n}{7}}-1} + \left(\frac{7}{6}\right)^{-\frac{n}{12}} + \left(\frac{7}{6}\right)^{-\frac{n}{13}}.
\end{eqnarray}
This concludes the proof of the Theorem. \qed

\medskip

We note here that the upper bound on $\epsilon$ in the statement of Theorem \ref{theorem-clique} is only used to facilitate exposition of our arguments in the proof. We claim that this upper bound can be increased further by using the same proof ideas, but that we cannot reach arbitrarily close to $\frac{1}{2}$. However, we conjecture that the constant $\epsilon$ in Theorem \ref{theorem-clique} can be as close to $\frac{1}{2}$ as desired, as long as it remains bounded away from it.

\begin{conjecture}
Let $\epsilon$ be a positive constant strictly less than $\frac{1}{2}$. For large enough $n$, starting from $\epsilon n$ agents of type $r$ and $(1-\epsilon) n$ agents of type $g$ on the clique $K_n$, the probability that the clique eventually contains only agents of type $r$ is at most $e^{-\Theta(n)}$.
\end{conjecture}

We also conjecture the following lower bound:

\begin{conjecture}
Starting from a single agent of type $r$ and $n-1$ agents of type $g$ on the clique $K_n$, the probability that the clique eventually contains only agents of type $r$ is at least $e^{-\Theta(n)}$. 
\end{conjecture}

\subsection{Minority domination} \label{Section:Minority domination}

\begin{figure}[htb]
\centering 
\includegraphics[scale=0.7]{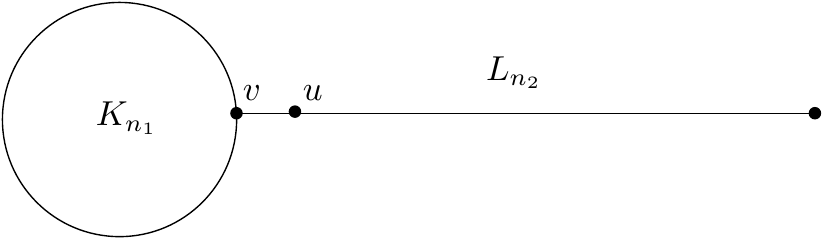} 
\caption{A lollipop graph consisting of a $K_{n_1}$ clique and a $L_{n_2}$ line.}
\label{lollipop}
\end{figure}

Consider the lollipop graph (see Figure \ref{lollipop}), which consists of a complete graph $K_{n_1}$ on $n_1$ vertices, among which vertex $v$ is connected to the leftmost vertex $u$ of a line graph $L_{n_2}$ on $n_2$ vertices, with $n_1+n_2=n$. Suppose that initially vertex $v$ and all vertices in $L_{n_2}$ are of type $r$, while all vertices in $K_{n_1} \backslash \{v\}$ are of type $g$. Without loss of generality, we also assume that $n_1 < n_2$, so that the color green is the initial minority in the graph. In this subsection we will provide a lower bound on the probability that the initial minority eventually wins, given that we start with this configuration. 

For the analysis we will use the analysis of Subsection \ref{Section:Clique}, together with a suitable domination argument and the following fact:

\begin{lemma} \label{Lemma:Path}
Consider a line graph $L_{m}$, in which the leftmost vertex is of type $g$ and all other $m-1$ vertices are of type $r$. Starting from this initial configuration, the probability that eventually all vertices become of type $g$ is $\frac{1}{2(m-1)}$.  
\end{lemma}
\proof Let ${\cal M}$ denote the Markov chain that describes the evolution of the protocol. Without loss of generality, we will assume that only transitions that result in some vertex changing its type are allowed. In particular, discarding any transitions of ${\cal M}$ resulting from interactions of the form $b \to r, b \to g, b \to b, g \to g$ or $r \to r$ does not change the probability of absorption of ${\cal M}$ to one of its two absorbing states (namely the one with all vertices of type $r$ and the one with all vertices of type $g$). 

Let us now construct a stochastic process ${\cal B}$ by taking snapshots of ${\cal M}$ every two transitions that change the state of ${\cal M}$. Then we can describe the set of states reachable by ${\cal M}$ (given the initial configuration) by a single number between $0$ and $m$. Indeed, starting from a single vertex of type $g$ on the leftmost vertex of the line and all other vertices of type $r$, and taking snapshots of ${\cal M}$ every two transitions, we can only reach configurations $S_k$, $k=0, \ldots, m$, in which a number of $k$ consecutive vertices to the left are of type $g$ and all others are of type $r$. By the Markov property we can easily verify that ${\cal B}$ is a Markov chain with state space $\{S_0, S_1, \ldots, S_m\}$ (where $S_0, S_m$ are absorbing) and transition probability matrix $P$ given by

\begin{equation}
P(S_i, S_j) = \left\{ 
\begin{array}{ll}
	\frac{1}{4}, & \quad \textrm{if $|j-i|=1$, for $i=2, 3, \ldots, m-2$,} \\
	\frac{1}{4}, & \quad \textrm{if $j=i+1=2$ or $j=i-1=m-2$,} \\
	\frac{1}{2}, & \quad \textrm{if $j=i+1=m$ or $j=i-1=0$,} \\
	1 - P(S_i, S_{i+1}) - P(S_i, S_{i-1}), & \quad \textrm{if $i=j$, for $i = 1, 2, \ldots, m-1$,} \\
	1, & \quad \textrm{if $i=j=0$ or $i=j=m$}.
\end{array}
\right.
\end{equation} 
In particular, we have that

\begin{equation}
\frac{P(S_i, S_{i-1})}{P(S_i, S_{i+1})} = 1
\end{equation}
for all $i=2, 3, \ldots, m-2$ and $\frac{P(S_1, S_0)}{P(S_1, S_2)} = \frac{P(S_{m-1}, S_{m})}{P(S_{m-1}, S_{m-2})} = 2$. Furthermore, we have that

\begin{equation}
\Pr\{\textrm{${\cal M}$ reaches $S_m$}| {\cal M}(0)=S_1\} = \Pr\{\textrm{${\cal B}$ reaches $S_m$}| {\cal B}(0)=S_1\}.
\end{equation}
In order to compute the above probability, we proceed as in the proof of Lemma \ref{Birth-death-absorption}. In particular, let $h_i = \Pr\{\textrm{absorption at $S_m$}| {\cal B}(0)=S_i\}$. We then have that $h_0 = 1 - h_m = 0$ and $h_{i+1} - h_i = \frac{P(S_i, S_{i-1})}{P(S_i, S_{i+1})} (h_i-h_{i-1})$, for $i = 1, \ldots, m-1$. In particular, $h_{i+1} - h_i = 2 h_1$, for all $i = 1, \ldots, m-2$ and $h_m-h_{m-1} = h_1$. Finally, we have that $1 = \sum_{i=0}^{m-1} (h_{i+1} - h_i) = 2 (m-1) h_1$, which completes the proof of the Lemma. \qed

\medskip

For the proof of the main Theorem in this subsection, we define stochastic processes ${\cal W}'$ and ${\cal C}'$ just as the processes ${\cal W}$ and ${\cal C}$ from Subsection \ref{Section:Clique}, but concerning the clique $K_{n_1}$. In particular,these processes take into account only transitions involving either edges that belong to the clique or the directed edge $(u, v)$. We note that Lemma \ref{relation-lemma} continues to hold for the modified process ${\cal C}'$ with only one modification (because of the existence of the edge $\{v, u\}$) concerning its second part: $T$ must satisfy ${\cal C'}_{|T} = 0$ and ${\cal C'}_{|T-1} > 0$ in order to be able to deduce that $K_{n_1}$ has only vertices of type $g$. \footnote{If this is not the case, i.e.~if only ${\cal C'}_{|T} = 0$ is given, then we can still claim that all vertices except $v$ are of type $g$, whereas $v$ can be of type $g$ or $b$. However, this is not needed in our analysis.} Furthermore, because of the existence of the directed edge $(u, v)$, the domination Lemmas \ref{Blank-domination-lemma} and \ref{Contest-domination-lemma} become as follows:

\begin{lemma}[Domination of ${\cal W}'$] \label{New-Blank-domination-lemma}
Let $\alpha', \beta', \kappa' \in \{1, \ldots, n_1-1\}$, with $\alpha'< \beta'$. Let also ${\cal B}_{\cal W'}$ be a birth-death process, which has state space ${\cal S_{{\cal B}_{\cal W'}}} = \{S'_0, \ldots, S'_{n_1}\}$, with $S'_{n_1}$ an absorbing state and transition probability matrix $P'$, with $P'(S'_i, S'_{i+1})=1$ for all $i \in \{0, \ldots, \alpha'\} \cup \{\beta', \ldots, n_1-1\}$, $\frac{P'(S'_i, S'_{i+1})}{P'(S'_i, S'_{i-1})} = \frac{2\kappa'+1}{\alpha'}$ for all $i \in \{\alpha'+1, \ldots, \beta'-1\}$ and $P'(S'_i, S'_i) = \Pr({\cal W}'(t) = i| {\cal W}'(t-1) = i)$, for all $t \geq 1$ and for all $i \in \{\alpha'+1, \ldots, \beta'-1\}$. Then, given that the vertices of type $r$ in $K_{n_1}$ are at most $\kappa'$, the process ${\cal W}'$ is stochastically dominated by ${\cal B}_{\cal W'}$ in the following sense: $\Pr({\cal W}'(t) > x| {\cal W}'(0)=0) \leq \Pr({\cal B}_{\cal W'}(t) \in \cup_{y>x} S'_x| {\cal B}_{\cal W'}(0)=S'_0)$, for any time $t$ and $x \in \{0, \ldots, n_1\}$.
\end{lemma}
\proof It suffices to show that, for any $t \geq 1$ and for any $i \in \{0, \ldots, n_1-1\}$, 

\begin{equation}
\Pr({\cal W}'(t) = i+1| {\cal W}'(t-1)=i) \leq \Pr({\cal B}_{\cal W'}(t) = S'_{i+1}| {\cal B}_{\cal W'}(t-1) = S'_i).
\end{equation}
This is trivially true for all $i \in \{0, \ldots, \alpha'\} \cup \{\beta', \ldots, n_1-1\}$, because the right hand side of the above inequality is 1. For $i \in \{\alpha'+1, \ldots, \beta'-1\}$, given that ${\cal W}'(t-1)=i$ (i.e.~there are exactly $i$ blanks at time $t$), the probability that ${\cal W}'$ increases by 1 in the next time step is equal to the probability that either a transition $g \to r$ or a transition $r \to g$ occurs either involving vertices that are both inside the clique, or involving $u$ and $v$. If we denote by $R'_{t-1}$ (resp. $G'_{t-1}$) the set of vertices of type $r$ (resp. of type $g$) in the clique $K_{n_1}$ at time $t-1$, then the above probability is at most $\frac{2 |R'_{t-1}| |G'_{t-1}|+1}{n_1(n_1-1) +1}$. On the other hand, the probability that ${\cal W}'$ decreases by 1 in the next time step is equal to the probability that either a transition $g \to b$ or a transition $r \to b$ occurs either involving vertices that are both inside the clique, or involving the edge $(u, v)$. Given that ${\cal W}'(t-1)=i$, this probability is at least $\frac{i(|R'_{t-1}| + |G'_{t-1}|)}{n_1(n_1-1)+1}$. But then 

\begin{equation}
\frac{\Pr({\cal W}'(t) = i+1| {\cal W}'(t-1)=i)}{\Pr({\cal W}'(t) = i-1| {\cal W}'(t-1)=i)} \leq \frac{2|R'_{t-1}| |G'_{t-1}| +1}{i(|R'_{t-1}| + |G'_{t-1}|)} \leq \frac{2 |R'_{t-1}|+1}{i}.
\end{equation}
By assumption, this is at most $\frac{2 \kappa'+1}{\alpha',}$, which combined with the fact that $P'(S'_i, S'_i) = \Pr({\cal W}'(t) = i| {\cal W}'(t-1) = i)$, for all $i \in \{\alpha'+1, \ldots, \beta'-1\}$, concludes the proof of Lemma \ref{New-Blank-domination-lemma}. \qed

\begin{lemma}[Domination of ${\cal C}'$] \label{New-Contest-domination-lemma}
Let $\beta', \kappa'$ be positive integers, with $\beta'+\kappa'<n_1$. Let also ${\cal B}_{\cal C'}$ be a birth-death process, which has state space ${\cal S_{{\cal B}_{\cal C'}}} = \{T'_0, \ldots, T'_{n_1}\}$, with a reflecting barrier at $T'_0$, an absorbing state $T'_{n_1}$ and transition probability matrix $Q'$, with $Q'(T'_i, T'_{i+1}) = 1$ for all $i \in \{0\} \cup \{\kappa', \ldots, n_1-1\}$, $\frac{Q'(T'_i, T'_{i+1})}{Q'(T'_i, T'_{i-1})} = \left(1 + \frac{1}{n_1-\beta'-\kappa'}\right) \frac{\kappa'+1}{n_1-\beta'-\kappa'}$ for all $i \in \{1, \ldots, \kappa'-1\}$ and $Q'(T'_i, T'_i) = \Pr({\cal C}'(\tau) = i| {\cal C}'(\tau-1) = i)$, for all $\tau \geq 1$ and for all $i \in \{1, \ldots, \kappa'-1\}$. Then, given that the vertices of type $b$ are at most $\beta'$, the process ${\cal C}'$ is stochastically dominated by ${\cal B}_{\cal C'}$ in the following sense: $\Pr({\cal C}'(\tau) > x| {\cal C}'(0)=|R'_0|) \leq \Pr({\cal B}_{\cal C'}(\tau) \in \cup_{y>x} T'_x| {\cal B}_{\cal C'}(0)=|R'_0|)$, for any time $\tau$ and $x \in \{0, \ldots, n_1\}$, where $|R'_0|$ is the number of vertices of type $r$ in $K_{n_1}$ at time 0.
\end{lemma}
\proof It suffices to show that, for any $\tau \geq 1$ and for any $i \in \{1, \ldots, n_1-1\}$, 

\begin{equation}
\Pr({\cal C'}(\tau) = i+1| {\cal C'}(\tau-1)=i) \leq \Pr({\cal B}_{\cal C'}(\tau) = T'_{i+1}| {\cal B}_{\cal C'}(\tau-1) = T'_i).
\end{equation}
This is trivially true for all $i \in \{0\} \cup \{\kappa', \ldots, n_1-1\}$, because the right hand side of the above inequality is 1. For $i \in \{1, \ldots, \kappa'-1\}$, we apply the \emph{principle of deferred decisions}. In particular, let $t_1$ (resp. $t_2$) be the time in the time scale of ${\cal M}$ (i.e.~the Markov chain describing the evolution of the protocol on $H$) that corresponds to the first (resp. second) transition of the $\tau$-th transition pair in the definition of ${\cal C}'$. Given that ${\cal C}'(\tau-1)=i$, the probability that ${\cal C}'$ increases by 1 in the next time step in the time scale of ${\cal C}'$ is equal to the probability that the $\tau$-th transition pair is $(r \to g, r \to b)$. If we denote by $R'_t$ (resp. $G'_t$) the set of vertices of type $r$ (resp. of type $g$) in the clique $K_{n_1}$ at time $t$, then the above probability is at most $\frac{|R'_{t_1}||G'_{t_1}|+1}{2|R'_{t_1}||G'_{t_1}|+1} \frac{|R'_{t_2}| (n_1-|R'_{t_2}|-|G'_{t_2}|)+1}{(|R'_{t_2}|+|G'_{t_2}|) (n_1-|R'_{t_2}|-|G'_{t_2}|)+1}$. On the other hand, given that ${\cal C}'(\tau-1)=i$, the probability that ${\cal C}'$ decreases by 1 in the next time step is equal to the probability that the $\tau$-th transition pair in the definition of ${\cal C}'$ is $(g \to r, g \to b)$, which is at least $\frac{|G'_{t_1}||R'_{t_1}|}{2|R'_{t_1}||G'_{t_1}|+1} \frac{|G'_{t_2}| (n_1-|R'_{t_2}|-|G'_{t_2}|)}{(|R'_{t_2}|+|G'_{t_2}|) (n_1-|R'_{t_2}|-|G'_{t_2}|)+1}$. But then 

\begin{eqnarray}
\frac{\Pr({\cal C}'(\tau) = i+1| {\cal C}'(\tau-1)=i, t_1, t_2)}{\Pr({\cal C}'(\tau) = i-1| {\cal C}'(\tau-1)=i, t_1, t_2)} & \leq & \frac{|R'_{t_1}||G'_{t_1}|+1}{|R'_{t_1}||G'_{t_1}|} \frac{|R'_{t_2}| (n_1-|R'_{t_2}|-|G'_{t_2}|)+1}{|G'_{t_2}| (n_1-|R'_{t_2}|-|G'_{t_2}|)} \label{division-by-0} \\
& \leq & \left( 1 + \frac{1}{|G'_{t_1}|}\right) \left( \frac{|R'_{t_2}|}{|G'_{t_2}|} + \frac{1}{|G'_{t_2}|}\right).
\end{eqnarray}
We note here that we do not need to worry about division with 0 in the above fractions. In particular, by the assumption $\beta'+\kappa'<n_1$, we have that $|G'_{t_1}| |G'_{t_2}| \neq 0$. Furthermore, by the definition of the time $t_2$, there must exist at least one vertex of type $b$ at time $t_2$, which implies that $n_1-|R'_{t_2}|-|G'_{t_2}|>0$. Finally, we do not need to worry about $|R'_{t_1}| = 0$ for the following reason: by the definition of ${\cal C'}$, having no vertices of type $r$ and ${\cal C}(\tau-1) \geq 1$ at the same time would mean that there is a transition $g \to r$ that needs to be paired before any other transition of the form $r \to g$ and thus the fraction in the left hand side of (\ref{division-by-0}) would be 0 (which is smaller than the bound given in the statement of the Lemma).

Since $i \leq \kappa'-1$ by assumption, it follows by the first part of Lemma \ref{relation-lemma} that $|R'_{t_2}| < \kappa'$. Therefore, we have that $\frac{\Pr({\cal C}'(\tau) = i+1| {\cal C}'(\tau-1)=i)}{\Pr({\cal C}'(\tau) = i-1| {\cal C}'(\tau-1)=i)} \leq \left(1 + \frac{1}{n_1-\beta'-\kappa'}\right) \frac{\kappa'+1}{n_1-\beta'-\kappa'}$, which combined with the fact that $Q'(T'_i, T'_i) = \Pr({\cal C}'(\tau) = i| {\cal C}'(\tau-1) = i)$, for all $\tau \geq 1$ and for all $i \in \{1, \ldots, \kappa'-1\}$, concludes the proof of Lemma \ref{New-Contest-domination-lemma}. \qed

\medskip

We are now ready to prove the main result of this subsection.

\begin{theorem} \label{theorem-minority-domination}
Consider a lollipop graph on $n$ vertices, which consists of a complete graph $K_{n_1}$ on $n_1 \geq 100 \ln{n}$ vertices, among which vertex $v$ is connected to the leftmost vertex $u$ of a line graph $L_{n_2}$ on $n_2 = n-n_1$ vertices. Suppose that initially vertex $v$ and all vertices in $L_{n_2}$ are of type $r$, while all vertices in $K_{n_1} \backslash \{v\}$ are of type $g$. Then with high probability, eventually only vertices of type $g$ will remain in the graph. 
\end{theorem}
\proof For the proof, we modify accordingly the proof of the main Theorem in Subsection \ref{Section:Clique}. We begin by setting $\alpha' = \frac{n_1}{3}, \beta' = \frac{n_1}{2}$ and $\kappa' = \frac{n_1}{7}$. In that case, the forward bias of the birth-death process ${\cal B}_{\cal W'}$ at states $\{S'_{\alpha'+1}, S'_{\alpha'+2} \ldots, S'_{\beta'-1}\}$ is at most $r_{\cal W'} \stackrel{def}{=} \frac{6.1}{7}$ (taking also into account that $n_1 \to \infty$) and the forward bias of the birth-death process ${\cal B}_{\cal C'}$ at states $\{T'_1, T'_2, \ldots, T'_{\kappa'-1}\}$ is at most $r_{\cal C'} \stackrel{def}{=} \frac{2.1}{5}$.

We will also denote by $R'_t$ (resp. $G'_t$) the set of vertices of type $r$ (resp. of type $g$) in the clique $K_{n_1}$ at time $t$. For simplicity, the initial configuration will be denoted by ${\cal E}$.

We first provide a lower bound on the probability that the clique $K_{n_1}$ reaches a configuration where all its vertices (i.e.~including $v$) are of type $g$. To this end, we define the following events, which are similar to the events $A_1$ and $A_2$ used in the proof of the main Theorem in Subsection \ref{Section:Clique} (only now they are defined for the clique $K_{n_1}$):

\begin{description}

\item[(i)] $\Gamma_1$ is the event that ${\cal C}'$ reaches $\kappa$ before the number of vertices of type $b$ reaches $\beta'$.

\item[(ii)] $\Gamma_2$ is the event that the number of vertices of type $b$ reaches $\beta'$ before ${\cal C}'$ reaches either 0 or $\kappa'$.

\end{description}

Notice now that, by the second part of Lemma \ref{relation-lemma} (see also the discussion before Lemma \ref{New-Blank-domination-lemma} in this subsection), whenever we have ${\cal C}'_{|t} = 0$ and ${\cal C}'_{|t-1} > 0$, all vertices in $K_{n_1}$ must be of type $g$. In view of this, we are interested in lower bounding the probability that ${\cal C}'$ reaches 0 before either $\Gamma_1$ or $\Gamma_2$ happens, given that we initially start with a configuration where all vertices in $K_{n_1} \backslash \{v\}$ are of type $g$ and $v$ is of type $r$. Equivalently, we are interested in upper bounding the probability that $\Gamma_1 \cup \Gamma_2$ happen before ${\cal C}'$ reaches 0. But for the latter, we can follow the same steps as in the proof of Theorem \ref{theorem-clique}. In particular, using $n_1, \alpha', \beta', \kappa', r_{\cal W'}, r_{\cal C'}$, instead of $n, \alpha, \beta, \kappa, r_{\cal W}, r_{\cal C}$ respectively, setting $k'_0 = k'_1 = \left( \frac{1}{r_{\cal W'}}\right)^{\frac{n_1}{12}}$, instead of $k_0, k_1$, and setting $\epsilon = \frac{1}{n_1}$ (i.e.~initially we start with a single vertex of type $r$ in $K_{n_1}$) we get the following equivalent to equation (\ref{fin-eq}):

\begin{eqnarray}
\Pr\{\textrm{$\Gamma_1 \cup \Gamma_2$ happens before ${\cal C}'$ reaches 0}| {\cal E}\} & \leq &\frac{\frac{1}{r_{\cal C'}}-1}{\left(\frac{1}{r_{\cal C'}}\right)^{\frac{n_1}{7}}-1} + k'_0 \left(\frac{1}{r_{\cal W'}}\right)^{-\frac{n_1}{6}} + \frac{O(n_1^5)}{k'_1} \\
& \leq & \left(\frac{7}{6.1}\right)^{-\frac{n_1}{13}} + \left(\frac{7}{6.1}\right)^{-\frac{n_1}{12}} + \frac{2.9}{2.1} \frac{1}{\left(\frac{5}{2.1}\right)^{\frac{n_1}{7}}-1} \\
&\leq & 8 \left(\frac{7}{6.1}\right)^{-\frac{n_1}{13}} \stackrel{def}{=} \phi_K.
\end{eqnarray}

By Lemma \ref{Lemma:Path}, we also have that, given that there are no vertices of type $r$ in the clique $K_{n_1}$ and that $v$ is of type $g$, the probability that eventually only vertices of type $g$ remain on the line $L_{n_2}$ is at least $\frac{1}{2n} \stackrel{def}{=} \phi_{L}$. Indeed, this follows from the observation that $n \geq n_2$ and that the existence of any other vertex of type $g$ or $b$ inside the line can only increase the probability that the type of the leftmost vertex wins without $v$ ever changing its type from $g$ to $b$.  

Using all the above we then have that, given ${\cal E}$, the probability that eventually only vertices of type $r$ remain in the lollipop graph at most the probability that at some point $\Gamma_1 \cup \Gamma_2$ happens before ${\cal C}'$ reaches 0, which is at most 

\begin{eqnarray}
\sum_{i=0}^{\infty} \phi_{K} (1 - \phi_L)^i = \frac{\phi_{K}}{\phi_L} = 16n \left(\frac{7}{6.1}\right)^{-\frac{n_1}{13}},
\end{eqnarray}
which is $o(1)$ (i.e.~it goes to 0 as $n$ goes to infinity) provided $n_1 \geq 100 \ln{n}$. This completes the proof of the Theorem. \qed

\subsection{Expected Exponential time to absorption} \label{Section:Exponential}

\begin{figure}[htb]
\centering 
\includegraphics[scale=0.7]{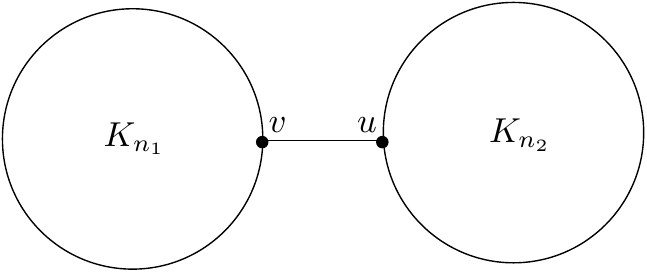} 
\caption{An interaction graph $G$ with 2 disjoint $K_{n_1}$ and $K_{n_2}$, connected with a single edge $u, v$.}
\label{exponentialexample}
\end{figure}

In this subsection we show that there are graphs in which the majority protocol of~\cite{AAE08} needs an expected exponential time to reach consensus. We prove this by analyzing the expected time to reach consensus in the case where the interaction graph consists of 2 disjoint cliques on $n_1$ and $n_2$ vertices each (i.e.~$n = n_1+n_2$) and a single edge between them, as shown in Figure \ref{exponentialexample}. Initially, all vertices in the $K_{n_1}$ clique are of type $g$, while all vertices in the $K_{n_2}$ clique are of type $r$.

We will assume w.l.o.g. that $n_1 \leq n_2$ and that $n_1$ is an increasing function of the total number of vertices (i.e.~$n_1 \to \infty$ as $n$ goes to infinity). In order to provide a lower bound on the time after which only one type of vertices remains (e.g. either $r$ or $g$), we will use a directed graph $H$ consisting of a single clique $K_{n_1}$ and a vertex $v$ outside the clique, which is connected to a vertex $u$ of the clique with a directed edge $(v, u)$.

\begin{figure}[htb]
\centering 
\includegraphics[scale=0.7]{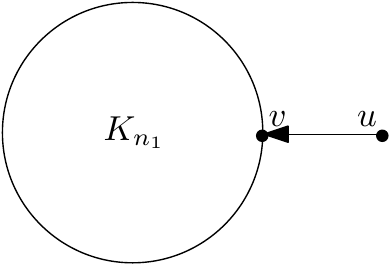} 
\caption{The graph $H$ consisting of a clique $K_{n_1}$ with a directed edge $(u, v)$, where only $u$ does not belong to the clique.}
\label{reducedexponentialexample}
\end{figure}

We now prove the following:

\begin{lemma} \label{directed-intermediate-lemma}
Let $T_H$ be the number of steps that the majority protocol of~\cite{AAE08} needs to reach consensus when the underlying graph is $H$ and initially all vertices are of type $g$ except for $u$ which is of type $r$. Then the mean value of $T_H$ is at least exponential in $n_1$.
\end{lemma}
\proof It is evident from the definition of the protocol of~\cite{AAE08} that eventually only vertices of type $r$ will remain. Indeed, vertex $u$ can never change its type (notice that the protocol of Angluin et al. is not symmetric), but it can affect the type of $v$, and through $v$ the type of every vertex in the clique. In order to analyze the time needed to reach (forced) consensus, we define stochastic processes ${\cal W}''$ and ${\cal C}''$ just as the processes ${\cal W}$ and ${\cal C}$ from Subsection \ref{Section:Clique}, but concerning the clique $K_{n_1}$. In particular,these processes take into account only transitions involving either edges that belong to the clique or the directed edge $(u, v)$. We note that the first part of Lemma \ref{relation-lemma} continues to hold for process ${\cal C}''$. However, because of the existence of the directed edge $(u, v)$, the domination Lemmas \ref{Blank-domination-lemma} and \ref{Contest-domination-lemma} become as the next two Lemmas (their proofs are identical to the proofs of Lemmas \ref{New-Blank-domination-lemma} and \ref{New-Contest-domination-lemma} respectively).

\begin{lemma}[Domination of ${\cal W}''$] \label{Double-New-Blank-domination-lemma}
Let $\alpha'', \beta'', \kappa'' \in \{1, \ldots, n_1-1\}$, with $\alpha''< \beta''$. Let also ${\cal B}_{\cal W''}$ be a birth-death process, which has state space ${\cal S_{{\cal B}_{\cal W''}}} = \{S''_0, \ldots, S''_{n_1}\}$, with $S''_{n_1}$ an absorbing state and transition probability matrix $P''$, with $P''(S''_i, S''_{i+1})=1$ for all $i \in \{0, \ldots, \alpha''\} \cup \{\beta'', \ldots, n_1-1\}$, $\frac{P''(S''_i, S''_{i+1})}{P''(S''_i, S''_{i-1})} = \frac{2\kappa''+1}{\alpha''}$ for all $i \in \{\alpha''+1, \ldots, \beta''-1\}$ and $P''(S''_i, S''_i) = \Pr({\cal W}''(t) = i| {\cal W}''(t-1) = i)$, for all $t \geq 1$ and for all $i \in \{\alpha''+1, \ldots, \beta''-1\}$. Then, given that the vertices of type $r$ in $K_{n_1}$ are at most $\kappa''$, the process ${\cal W}''$ is stochastically dominated by ${\cal B}_{\cal W''}$ in the following sense: $\Pr({\cal W}''(t) > x| {\cal W}''(0)=0) \leq \Pr({\cal B}_{\cal W''}(t) \in \cup_{y>x} S''_y| {\cal B}_{\cal W''}(0)=S''_0)$, for any time $t$ and $x \in \{0, \ldots, n_1\}$.
\end{lemma}

\begin{lemma}[Domination of ${\cal C}''$] \label{Double-New-Contest-domination-lemma}
Let $\beta'', \kappa''$ be positive integers, with $\beta''+\kappa''<n_1$. Let also ${\cal B}_{\cal C''}$ be a birth-death process, which has state space ${\cal S_{{\cal B}_{\cal C''}}} = \{T''_0, \ldots, T''_{n_1}\}$, with a reflecting barrier at $T''_0$, an absorbing state $T''_{n_1}$ and transition probability matrix $Q''$, with $Q''(T''_i, T''_{i+1}) = 1$ for all $i \in \{0\} \cup \{\kappa'', \ldots, n_1-1\}$, $\frac{Q''(T''_i, T''_{i+1})}{Q''(T''_i, T''_{i-1})} = \left(1 + \frac{1}{n_1-\beta''-\kappa''}\right) \frac{\kappa''+1}{n_1-\beta''-\kappa''}$ for all $i \in \{1, \ldots, \kappa''-1\}$ and $Q''(T''_i, T''_i) = \Pr({\cal C}''(\tau) = i| {\cal C}''(\tau-1) = i)$, for all $\tau \geq 1$ and for all $i \in \{1, \ldots, \kappa''-1\}$. Then, given that the vertices of type $b$ are at most $\beta''$, the process ${\cal C}''$ is stochastically dominated by ${\cal B}_{\cal C''}$ in the following sense: $\Pr({\cal C}''(\tau) > x| {\cal C}''(0)=|R''_0|) \leq \Pr({\cal B}_{\cal C''}(\tau) \in \cup_{y>x} T''_x| {\cal B}_{\cal C''}(0)=|R''_0|)$, for any time $\tau$ and $x \in \{0, \ldots, n_1\}$, where $|R''_0|$ is the number of vertices of type $r$ in $K_{n_1}$ at time 0.
\end{lemma}

Continuing with the proof of Lemma \ref{directed-intermediate-lemma}, we note that the expected time needed for the protocol to reach consensus (i.e.~reach a configuration where all vertices are of type $r$) is at least the time that it needs to reach either $\beta''$ vertices of type $b$ or $\kappa''$vertices of type $r$.

We now set $\alpha'' = \frac{n_1}{3}$, $\beta'' = \frac{n_1}{2}$ and $\kappa'' = \frac{n_1}{7}$. In that case, the forward bias of the birth-death process ${\cal B}_{\cal W''}$ at states $\{S''_{\alpha''+1}, S''_{\alpha''+2} \ldots, S''_{\beta''-1}\}$ is at most $r_{\cal W''} \stackrel{def}{=} \frac{6.1}{7} <1$ (taking also into account that $n_1 \to \infty$) and the forward bias of the birth-death process ${\cal B}_{\cal C''}$ at states $\{T''_1, T''_2, \ldots, T''_{\kappa''-1}\}$ is at most $r_{\cal C''} \stackrel{def}{=} \frac{2.1}{5}<1$. 

In particular, by Lemma \ref{Birth-death-Time}(b) and by the domination Lemma \ref{Double-New-Blank-domination-lemma} for process ${\cal W}''$, the expected number of steps needed in order to reach $\beta''$ vertices of type $b$ (given that initially there are no vertices of type $b$) is $\Omega\left( \frac{1}{\frac{1}{r_{\cal W''}} -1}\left( \frac{1}{r_{\cal W''}}\right)^{\frac{n_1}{6}}\right) = \Omega\left(\left( \frac{7}{6.1}\right)^{\frac{n_1}{6}}\right)$. Similarly, by Lemma \ref{Birth-death-Time}(b) and by the domination Lemma \ref{Double-New-Contest-domination-lemma} for process ${\cal C}''$, the expected number of steps in the time scale of ${\cal C}''$ needed for ${\cal C}''$ to reach $\kappa''$ (given that initially only vertex $u$ is of type $r$) is at least $\Omega\left(\frac{1}{\frac{1}{r_{\cal C''}} -1}\left( \frac{1}{r_{\cal C''}}\right)^{\frac{n_1}{7}}\right) = \Omega\left(\left( \frac{5}{2.1}\right)^{\frac{n_1}{7}}\right)$. By the first part of Lemma \ref{relation-lemma}, this is at least the expected time needed for the number of vertices of type $r$ to reach $\kappa''$. The proof of Lemma \ref{directed-intermediate-lemma} is completed by noting that the expected time needed for the protocol to reach the configuration where are vertices are of type $r$ is at least the minimum of these two exponential quantities. \qed

\medskip

Using Lemma \ref{directed-intermediate-lemma}, we can prove the main result of this subsection.

\begin{theorem}
Let $G$ be an interaction graph on $n$ vertices, which consists of 2 disjoint cliques on $n_1$ and $n_2$ vertices each and a single edge between them (see Figure \ref{exponentialexample}). Suppose that initially, all vertices in the $K_{n_1}$ clique are of type $g$, while all vertices in the $K_{n_2}$ clique are of type $r$. Then the majority protocol of~\cite{AAE08} needs an expected exponential number of steps to reach consensus. 
\end{theorem}
\proof Let $T_G$ be the number of steps needed for the protocol to reach consensus, starting from the intial configuration described in the Theorem. Let also $A$ be the event that the initial majority eventually wins. We then have that

\begin{equation}
\mathbb{E}[T_G] = \Pr(A) \mathbb{E}[T_G|A] + (1 - \Pr(A)) \mathbb{E}[T_G|\overline{A}] \geq \Pr(A) \mathbb{E}[T_G|A].
\end{equation}

Assume now w.l.o.g. that $n_1 \leq n_2$. Then, by a similar argument to the derivation of equation (\ref{dominationeq}) in the proof of Theorem \ref{highermajority}, the probability that the initial majority wins (i.e.~eventually only vertices of type $r$ remain), is at least $\frac{1}{2}$. The proof is concluded by noting that if $T_H$ is defined as in Lemma \ref{directed-intermediate-lemma} then $\mathbb{E}[T_G|A] \geq \mathbb{E}[T_H]$. \qed



{
\bibliographystyle{abbrv}
\bibliography{majority-ref}
}

\end{document}